\documentclass[reprint,amsmath,amssymb,aps]{revtex4-1}

\usepackage{units}
\usepackage{ulem}
\usepackage{amsmath}
\usepackage{amssymb}
\usepackage{amsfonts}
\usepackage{epstopdf}
\usepackage{microtype}
\usepackage{graphicx} 			
\usepackage{bm}				
\usepackage{color}			
\usepackage{pstricks}			
\usepackage{float}			

\usepackage{soul}

    \newcommand{\be}		{\begin{equation}}	 	
    \newcommand{\ee}		{\end{equation}}	 	
    \newcommand{\mcal}[1]	{\mathcal{#1}}		 	

    \newcommand{\Eq}[1]		{Eq.~(\ref{#1})}		
    \newcommand{\Fig}[1]	{Fig.~\ref{#1}}			
    	    
    \newcommand{\ex}[1]		{\mbox{e}^{#1}}			
    			
\def \hH{ \hat{\mathcal{H}}}

\newcommand{\PD}{\phantom{\dag}}
\newcommand{\ev}[1]{\ensuremath{\left\langle #1 \right\rangle}}
\newcommand{\evS}[1]{\ensuremath{ \langle #1  \rangle}}

\newcommand{\IN}{\textup{in}}
\newcommand{\OUT}{\textup{out}}

\newcommand{\HOP}{\textup{hop}}

\newcommand{\ADD}{\textup{add}}
\newcommand{\DRIVE}{\textup{drive}} 

\newcommand{\di}{\hat d_{i}}
\renewcommand{\dj}{\hat d_{j}}
\newcommand{\dij}{\hat d_{ij}}
\newcommand{\did}{\hat d_{i}^{\dag}} 
\newcommand{\djd}{\hat d_{j}^{\dag}}
\newcommand{\dijd}{\hat d_{ij}^{\dag}}
\newcommand{\kij}{\kappa_{ij}}

\begin{document}
\title{\textsf{Nonreciprocal Signal Routing in an Active Quantum Network}}
\author{\textsf{A.  Metelmann}}
\author{\textsf{H. E. T\"ureci}}
\affiliation{\textsf{Department of Electrical Engineering, Princeton University, Princeton, New Jersey 08544, USA}}
\date{\today}

\begin{abstract} 
As superconductor quantum technologies are moving towards large-scale integrated circuits, a robust and flexible approach to routing photons at the quantum level becomes a critical problem. Active circuits, which contain parametrically driven elements selectively embedded in the circuit offer a viable solution. Here, we present a general strategy for routing nonreciprocally quantum signals between two sites of a given lattice of oscillators, implementable with existing superconducting circuit components. Our approach makes use of a dual lattice of overdamped oscillators linking the nodes of the main lattice. Solutions for spatially selective driving of the lattice elements can be found, which optimally balance coherent and dissipative hopping of microwave photons to nonreciprocally route signals between two given nodes. In certain lattices these optimal solutions are obtained at the exceptional point of the dynamical matrix of the network. We also demonstrate that signal and noise transmission characteristics can be separately optimized. 
\end{abstract}
\maketitle

\section{Introduction}
%
In large-scale integrated electronic circuits, active components play an important role in isolating, routing and amplifying electronic signals. Active components rely on an external energy source that pumps energy into the circuit to control the transport of signals. The important role of active components as fundamental primitives for quantum state control and read-out in large-scale circuits is also being recognized in superconducting quantum technologies \cite{devoret_superconducting_2013}. In recent years, the use of parametric interactions have emerged as an effective and versatile method to implement such active components. In particular, efforts are underway to build chip-scale non-magnetic directional amplifiers and circulators based on parametric interactions  \cite{kamal_noiseless_2011, abdo_directional_2013, abdo_josephson_2014, kamal_asymmetric_2014, estep_magnetic-free_2014, ranzani_graph-based_2015, sliwa_reconfigurable_2015, kerckhoff_-chip_2015-1, lecocq_nonreciprocal_2016, bernier_nonreciprocal_2016, fang_generalized_2017}. 
These approaches rely on dynamic-modulation induced nonreciprocity in a circuit of a few oscillators. Similar chiral circuits are also being studied as potential building blocks for creating fractional Quantum Hall states of light \cite{roushan_chiral_2017}. 

Nonreciprocal active circuits rely on the general strategy of imparting a direction-dependent phase on the propagating microwave signal through the modulation of nonlinear elements with pre-specified phase relationships. This approach is analogous to the dynamic modulation of the refractive index in the optical domain \cite{hwang_all-fiber-optic_1997, yu_complete_2009, doerr_optical_2011, fang_photonic_2012, lira_electrically_2012, tzuang_non-reciprocal_2014}. In the optical domain the accompanying non-zero imaginary part of the refractive index and the associated noise limit the performance of these devices for the processing of quantum signals. In microwave circuits, the availability of lossless JJ-based superconducting circuit components allows the close-to-ideal implementation of an effective time-varying reactance 
\cite{yurke_observation_1989}.  

Demonstration of loss-less nonreciprocal devices with quantum-limited noise performance in superconducting microwave circuits have set the stage for more complex directional circuits for isolating, routing and switching signals at the quantum level. In an ideal setting, a single photon injected into a node of a network should be nonreciprocally transported to another node, while selectively experiencing gain. In principle, nonreciprocal transport in a lattice is possible through the engineering of a lattice of resonators where the band structure of the bulk exhibits non-trivial topological properties. A finite-size realization, by bulk-edge correspondence, can contain one-way edge states which break reciprocity  \cite{koch_time-reversal-symmetry_2010, nunnenkamp_synthetic_2011, fang_realizing_2012, mei_simulation_2015, ningyuan_time-_2015, hu_measurement_2015}. Crucially, in these implementations it is desirable to minimize dissipation.

 \begin{figure}  
 \centering\includegraphics[width=0.4\textwidth]{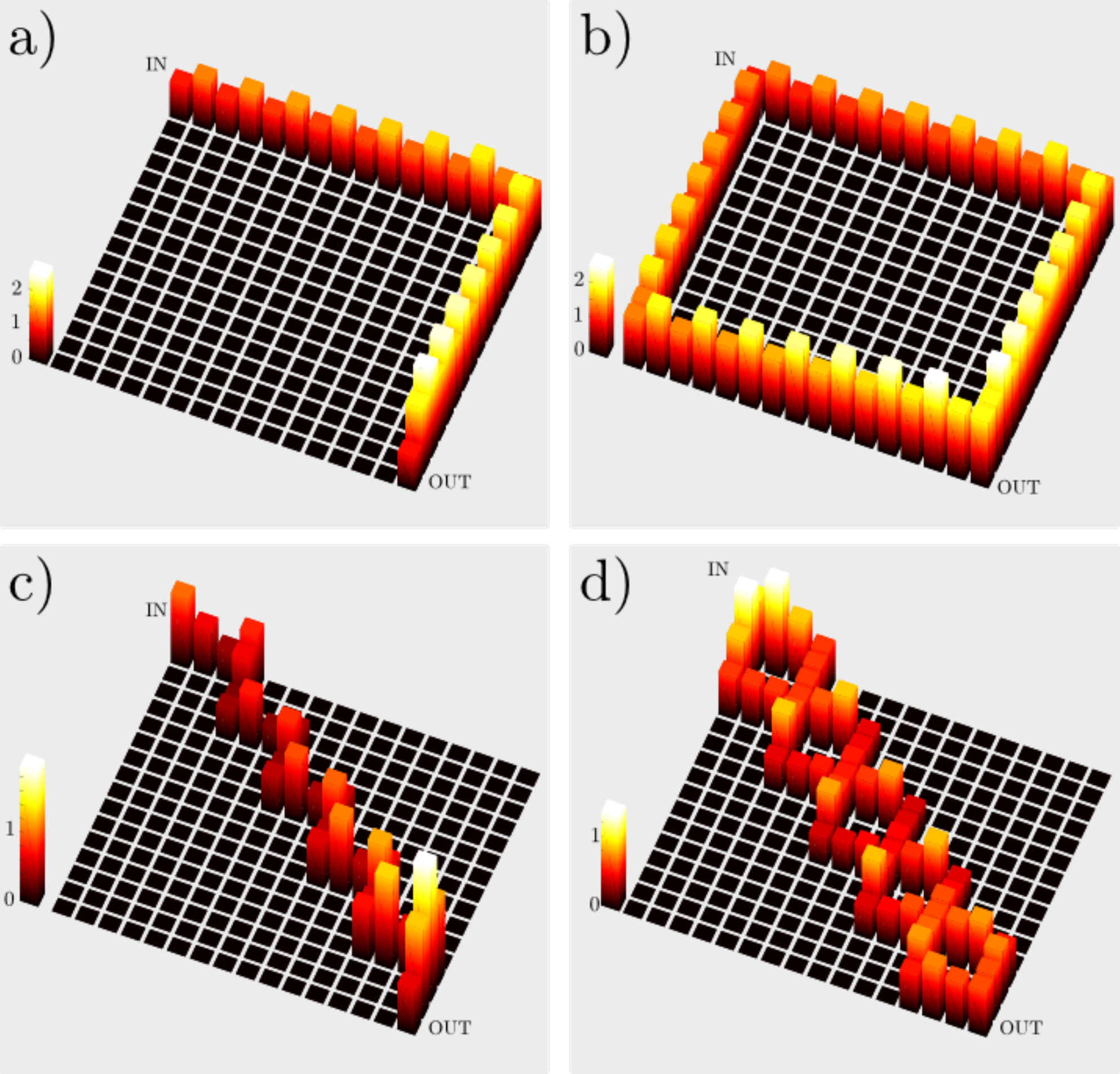}  
	\caption{Directional propagation in a oscillator lattice with 256 sites. 
 	         An input signal is injected on the upper left corner and propagates along a pre-designed path to the output waveguide attached to the lower right oscillator.
 	         Depicted is the averaged steady-state amplitude of each oscillator.}
	\label{fig:VariousPropagationWays}
 \end{figure}

Here we follow a different route that relies on dissipative stabilization to realize point-to-point nonreciprocal routing of photons in a lattice. Our approach makes use of a dual lattice of dissipative elements on the links connecting the nodes of the main lattice. These elements can be designed to implement a specific balance of unitary and dissipative hopping between two neighboring nodes \cite{metelmann_nonreciprocal_2015, metelmann_nonreciprocal_2017}. With respect to earlier approaches to active lattices \cite{fang_realizing_2012} dissipation here is deliberately designed to be comparable to the inter-site tunneling. Dissipative interactions can be designed to suppress all the routes other than the desired one, on which a nonreciprocal propagation takes place. We show further that the route can be switched dynamically by changing the spatial distribution of the modulation (its frequency, phase and amplitude) acting on the dual lattice (see \Fig{fig:VariousPropagationWays}). 
Moreover, we discuss amplification and associated noise characteristics of such an active lattice. We find that signal and noise propagating through the lattice can undergo different interference processes, which can be utilized to increase the signal-to-noise ratio. Finally, we present possible implementation schemes using existing superconducting circuit components.

\section{The Active Lattice Effective Hamiltonian}
\label{Sec:ActiveLattice}
\begin{figure}  
 \centering\includegraphics[width=0.45\textwidth]{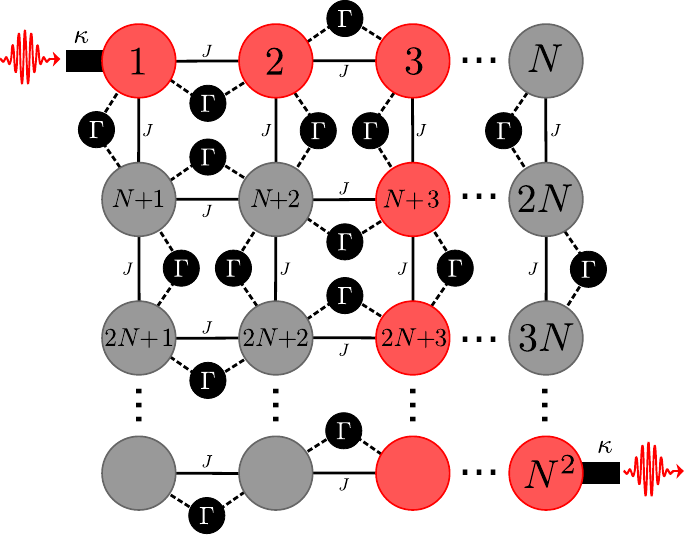} 
	\caption{Illustration of the active lattice.
	        Each oscillator of the $N^2$-lattice is directly coupled to its neighboring oscillators (solid lines). For simplicity we assume here uniform coupling strengths $G_{ij} e^{i\phi_{ij}} \equiv J$. 
	        Additionally, each oscillator pair is indirectly coupled  via link-oscillators (black circles), giving rise to an incoherent indirect exchange at the rate $\Gamma$. Oscillator 1 and $N^2$ are coupled to external waveguides with coupling strength $\kappa$.
	        The red circles denote a possible propagation path through the lattice if a signal is injected on oscillator 1
	        and transmitted to oscillator $N^2$.}
	\label{fig:SketchLattice}
 \end{figure}

%
We consider a lattice of oscillators (\Fig{fig:SketchLattice}) where the exchange of excitations between two nodes $d_i$ and $d_j$ takes place via two processes: a direct exchange (amplitude $G_{ij} \ex{ - i \phi_{ij}}$) and an indirect exchange via a link-oscillator $\dij$ (amplitudes $G_{i;ij}$, $G_{j;ij}$). The dynamics of such a system is governed by the effective Hamiltonian
\be \label{Eq.LatticeHam}
 \hH = \sum_{<i,j>}         
            G_{ij}  \;     \did  \dj  \ex{ - i \phi_{ij}}   
       +    G_{i;ij}  \;   \did  \dij
       +    G_{j;ij}  \;   \djd  \dij  + h.c.   .
\ee
Here, $<i,j>$ denotes nearest neighbor nodes and the indices $i$ and $j$ run over integers $1,\ldots, N^2$, from left to right and top to bottom. The hopping elements $G_{ij}$,  $G_{i;ij}$ and $G_{j;ij}$ are assumed to be real-valued. A crucial element here is the tunable non-zero phase $\phi_{ij}$. This lattice-model with adjustable parameters $G$ and $\phi$ can be realized through parametric processes, which will be discussed in Section~\ref{Sec.:Implementation}. We furthermore specify that each link-oscillator $\dij$ is coupled to a reservoir that gives rise to dissipation at rate $\kij$ and is subject to the associated noise. The goal is to design the parameters $G_{ij}$,  $G_{i;ij}$, $G_{j;ij}$, and $\phi_{ij}$ to nonreciprocally route an excitation injected from the site $i=1$ to the site $i=N^2$, where the signal is to be collected, as shown in \Fig{fig:SketchLattice}. 

We also consider amplification of the injected signal, which can be implemented by reconfiguring the parametric interactions on a given link oscillator, as discussed in Section~\ref{Sec.:Implementation}. This leads to the following interaction between the link and the node oscillators:
\be\label{Eq.PAHam}
 \hH_{\rm PA} =            
 \sum_{<i,j>}         
              G_{ij}    \;   \did  \dj   \ex{ - i \phi_{ij}}  
         +    G_{i;ij}  \;   \did  \dijd
         +    G_{j;ij}  \;   \djd  \dijd  + h.c.   ,
\ee
which can be optimized to yield one-way propagation with a tunable gain from oscillator $j$ to $j+1$.

 \begin{figure*}  
 \centering\includegraphics[width=1.0\textwidth]{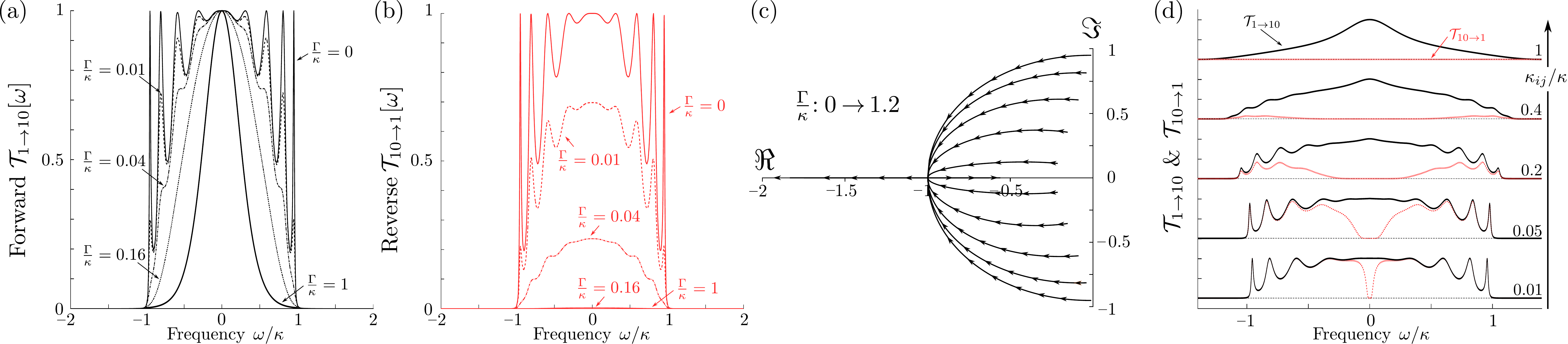}  
	\caption{Transmission properties and eigenvalues for a  chain of $N=10$ oscillators with fixed coherent hopping strength $|J| = \kappa/2$.
	        (a,b) Forward and backward transmission for various values of $\Gamma/\kappa$. The transmission window is determined by the coherent hopping strength, 
	        i.e., transmission in both directions is possible in the frequency range $\Delta\omega = 4|J|$. Once the directionality condition $\Gamma/\kappa = 1$ is met, 
	        the reverse transmission vanishes.
	        (c) Eigenvalues of the dynamical matrix for various values of $\Gamma/\kappa$. Below the exceptional point, i.e., $\Gamma/\kappa < 1$ the system shows underdamped oscillations.
	        The point of directionality at $\Gamma/\kappa = 1$ coincides with an exceptional point, where all eigenvalues are degenerate and real. 
	        For $\Gamma/\kappa > 1$ the eigenvalues are purely real, implying overdamped dynamics.
	        (d) Forward and backward transmission for fixed directionality condition, as the damping rate of the link oscillators $\kappa_{ij}$ is varied.
	        The bandwidth in which nonreciprocal transmission is possible is on the order of $2 \kappa_{ij}$. It is seen that the reverse transmission gets suppressed even for modest values of $\kappa_{ij}$.
	        }
	\label{fig:CavityChainProperties}
 \end{figure*}

\section{One-way transport between two nodes}

We first discuss the one of the basic ingredients of the proposed active lattice, namely the implementation of one-way transport between {\it two} isolated nodes, $i \rightarrow j$. We assume in what follows that $\kij$, the dissipation acting on the link oscillator $ij$, is the dominant loss channel in the system.  
Using a Heisenberg-Langevin approach \cite{SI}, the conditions for one-way propagation can be extracted by considering the dynamics of expectation values $\bar{d}_n \equiv \evS{ \hat d_n}$ and adiabatically eliminating the link-oscillator:
\begin{align}\label{Eq.EoMCouplingGeneral}
\dot{\bar{d}}_i = & - \frac{\Gamma_{i;ij}}{2}  \bar{d}_i -  \left[ i G_{ij}  \ex{-i\phi_{ij} }  + \frac{\sqrt{ \Gamma_{i;ij} \Gamma_{j;ij} }}{2}  \right]  \bar{d}_j ,
\nonumber \\
\dot{\bar{d}}_j  =&  - \frac{\Gamma_{j;ij}}{2} \bar{d}_j -  \left[ i  G_{ij} \ex{+i\phi_{ij} }  + \frac{\sqrt{ \Gamma_{i;ij} \Gamma_{j;ij} }}{2}  \right] \bar{d}_i .
 \end{align} 
Here $\Gamma_{n;ij} = 4 G_{n;ij}^2/ \kij, (n\in i,j)$. We aim for the situation where the oscillator $j$ is driven by the oscillator $i$ but not vice versa. This can be achieved through balancing the effective dissipative hopping term generated by the integration out of the link oscillator, the second term in the square brackets, with the unitary hopping term, given by the first term.  
The balancing conditions become
\be
\phi_{ij} = -\frac{\pi}{2} \quad \quad \text{and} \quad \quad  G_{ij} = \frac{\sqrt{ \Gamma_{i;ij} \Gamma_{j;ij} }}{2}.
\label{Eqnonrcond}
\ee
This condition provides a {\it manifestly} directional coupling for excitations (at the equation of motion level, $j$ is coupled to $i$ but not vice versa).
 
This mechanism for nonreciprocal transport between two oscillators through the balancing of dissipative and coherent hopping has been first proposed in Ref.~\cite{metelmann_nonreciprocal_2015}. Recent experimental work in three-mode systems have demonstrated superconducting circulators and directional amplifiers operating close to the standard quantum limit \cite{sliwa_reconfigurable_2015,lecocq_nonreciprocal_2016}. These experiments essentially implement conditions similar to the one stated in \Eq{Eqnonrcond}, as has been also found in Ref.~\cite{ranzani_graph-based_2015}. Nonreciprocal signal propagation via a dissipation-based approach was recently implemented in an optomechancial setup as well \cite{fang_generalized_2017}.   
The generalization of these considerations to an $N \times N$ lattice requires the satisfaction of further conditions while providing additional functionalities, which we discuss below. 
%
%
\section{Nonreciprocal Signal Propagation}
\label{Sec.:1Dsystem}
%
The evaluation of transport characteristics requires the system to be opened up to the environment. Besides that, we aim to design the finite dissipation on the link-oscillators to control the transport of excitations. Therefore the active lattice dynamics has to be addressed through an open system approach. This regime should be contrasted to earlier work in lattices subject to artificial gauge fields \cite{haldane_possible_2008, raghu_analogs_2008, hafezi_robust_2011,fang_realizing_2012,yannopapas_dirac_2013,lu_topological_2014,peano_topological_2015} where dissipation is generally expected to be minimized and does not play a critical role. In the latter case, directional propagation is possible due to topological protected edge states and can be described through Hamiltonian dynamics. Because the optimal conditions we find sensitively depend on the lattice dimensionality and geometry, we first discuss the case of a one-dimensional chain. 

We begin by illustrating the conditions for nonreciprocal transport in a chain of $N$ oscillators. Coupling the input ($d_{1}$) and output ($d_{N}$) oscillators to external waveguides, while giving rise to an adjustable coupling loss $\kappa$ (assumed equal for both ports without loss of generality), allows us to study signals entering and leaving the chain. For simplicity, we assume uniform couplings $G_{ij} e^{i\phi_{ij}} \equiv J$ and $\Gamma \equiv 4 G_{n;ij}/ \kij, (n\in i,j)$. The condition \Eq{Eqnonrcond} is then $J = - i \Gamma/2$, resulting in the decoupling of the $j$th oscillator from the  $j+1$th oscillator. This decoupling leads to a situation where an oscillator in the chain is driven by its left neighbor, but never from any higher element in the chain, i.e., the stationary solution of each oscillator becomes $\bar d_j = - \bar d_{j-1}$. 
By using standard input-output theory \cite{gardiner_quantum_2004}, $\hat d_{j,\OUT} = \hat d_{j, \IN} + \sqrt{\kappa} \hat d_j $, the scattering between the input and output ports is described by a $2 \times 2$ scattering matrix $\mathbf{s}[\omega]$:
\begin{align}\label{Eq:sDefn}
		\mathbf{D}_{\OUT}[\omega] = \mathbf{s}[\omega] \, \mathbf{D}_{\IN}[\omega] +  \vec{\hat{\xi}}[\omega],
		\hspace{0.2cm}
		\mathbf{D}[\omega]  =\left(  \hat d_{1 }[\omega] ,  \hat d_{N}[\omega] \right)^T. 
\end{align} 
Here, $\hat{\xi}[\omega]$ accounts for noise incident on the oscillators from the waveguides and the zero frequency scattering matrix is
\begin{align}\label{Eq.:SMatrixChain}
	\mathbf{s}[0] =  
	\left(
	\begin{array}{cc}
  		\displaystyle \frac{  \Gamma - \kappa }{\kappa + \Gamma}
		& 0 
	\\[2mm]
  		  \displaystyle \frac{ (-1 )^N 4 \kappa \Gamma}{(\kappa + \Gamma)^2}
		&\displaystyle \frac{  \Gamma - \kappa }{\kappa + \Gamma}
	\end{array}
	\right)
	\overset{\Gamma \equiv \kappa}{=}
		\left(
	\begin{array}{cc}
  		  0
		& 0 
	\\[2mm]
  		\left(-1\right)^N
		&0
	\end{array}
	\right).
\end{align}
Thus by applying the impedance matching condition $\Gamma = \kappa$ in the second step, we realize the scattering matrix of a perfect isolator. No input on oscillator $N$ will ever show up at the output of oscillator 1, while any input on oscillator 1 will be perfectly transmitted to oscillator $N$, i.e., $|s_{21}| = 1$. Interestingly, the impedance matching condition requires that $\kappa$, the coupling to input and output waveguides, to be the same order as the hopping strength $|J|$ ($=\Gamma/2$ to satisfy \Eq{Eqnonrcond}). We note that this condition is not necessary for nonreciprocity, but it prevents unwanted back-reflection of an injected signal.

Next we consider the transmission away from resonance. In the absence of incoherent hopping via the link oscillators, $\Gamma = 0$, forward and reverse transmission display $N$ resonance peaks and are identical as required by reciprocity [see~\Fig{fig:CavityChainProperties} (a,b) for $\Gamma/\kappa=0$]. We note that the operation point of choice requires that the chain is in the low-finesse regime ($|J| = \kappa/2$), thus not all resonances (in particular those near the center of the band) are well-resolved. As $\Gamma$ is turned on, forward transmission peaks in \Fig{fig:CavityChainProperties}(a) gradually smear out. When the directionality matching condition $\Gamma= 2|J| = \kappa$ is satisfied the entire band originally of width $4J$ collapses to a single Lorentzian peak with a bandwidth of the order of $\kappa/2$. Simultaneously reverse transmission [\Fig{fig:CavityChainProperties}(b)] is seen to vanish completely within the band.

The analysis of the spectrum of the dynamical matrix ${\mcal L}$, defined by $\dot{\bar{\bf{d}}} =  {\mcal L} \, \bar{\bf{d}}$ where $\bar{\bf{d}} = [  \evS{ \hat d_1} \evS{ \hat d_2} \cdots \evS{ \hat d_N}]^\text{T}$, reveals another interesting aspect of the nonreciprocity condition found. Due to the coupling to the waveguides and the dissipative link-oscillators, ${\mcal L}$ is a non-Hermitian matrix and its eigenvalues ${\mcal L}\,  \bf{v}_n = \varepsilon_n \bf{v}_n$ are generally complex-valued. For $\Gamma=0$, the system dynamics is governed by $N$ complex eigenvalues whose real and imaginary parts give the damping rates and the associated resonance frequencies respectively of Bloch modes of an open tight-binding oscillator chain (for vanishing waveguide coupling $\kappa \rightarrow 0$ the eigenvalues would be purely imaginary $\varepsilon_n = 2i |J| \cos[ n \pi/(N+1)]$ implying undamped, coherent dynamics). As $\Gamma$ approaches the nonreciprocity condition, the eigenvalues collapse to an N-fold degenerate {\it purely real} eigenvalue given by $\varepsilon_n = -\kappa$ [\Fig{fig:CavityChainProperties} (c)], implying overdamped dynamics. The inspection of eigenvectors reveals that all eigenvectors are degenerate as well, hence the nonreciprocity condition found coincides with an exceptional point. The role of such special degeneracies and their connection to unusual dynamical regimes have recently attracted a lot of interest in coupled optical cavities operating in the classical regime \cite{bender_observation_2013, peng_parity-time-symmetric_2014, chang_parity-time_2014, brandstetter_reversing_2014, hodaei_parity-timesymmetric_2014, peng_chiral_2016}.

A remaining important parameter is the damping rate $\kappa_{ij}$ of each link-oscillator, it determines the frequency band over which the transmission can be rendered nonreciprocal.
To sufficiently suppress the reverse transmission requires $\kappa_{ij}/\kappa > 1$ [\Fig{fig:CavityChainProperties}(d)].
The directionality bandwidth is on the order of $\Delta_d = 2\kappa_{ij}$, i.e., a detuning of $\Delta_d/2$ from resonance corresponds to a 3~dB isolation between forward and reverse transmission.

 \begin{figure}[t]  
 \centering\includegraphics[width=0.4\textwidth]{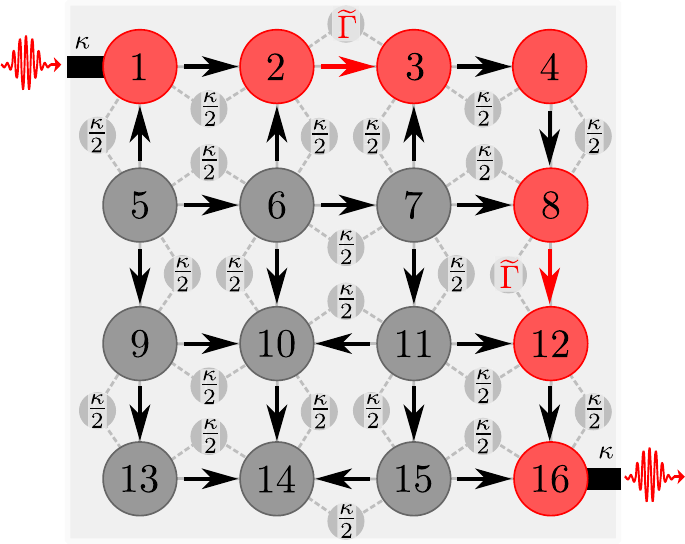} 
	\caption{Parametrization for nonreciprocal propagation along one edge in a 16 oscillator lattice.
	         We implement two amplification stages on the upper and the left edge with $\Gamma_{23} = \Gamma_{812}= \widetilde \Gamma$
	         and $\widetilde \phi_{23} = \widetilde \phi_{812} = \pi/2$. The remaining couplings are uniformly chosen to
	         be $\Gamma  = 2|J| = \kappa/2$. The phases depend on the propagation direction as indicated by the black arrows, we have
	         $\phi_{ij} = - \pi/2$ for $\rightarrow, \downarrow$  and $\phi_{ij} = + \pi/2$ for $\leftarrow, \uparrow$.
	         In principle, only the phases and couplings of the signal-carrying oscillators (red circles) and their nearest neighbors have to be fixed.
	         The remaining oscillators do not affect the transmission properties of the lattice.}
	\label{fig:SketchOneEdge}
 \end{figure}

 \begin{figure*} 
 \centering\includegraphics[width=1.0\textwidth]{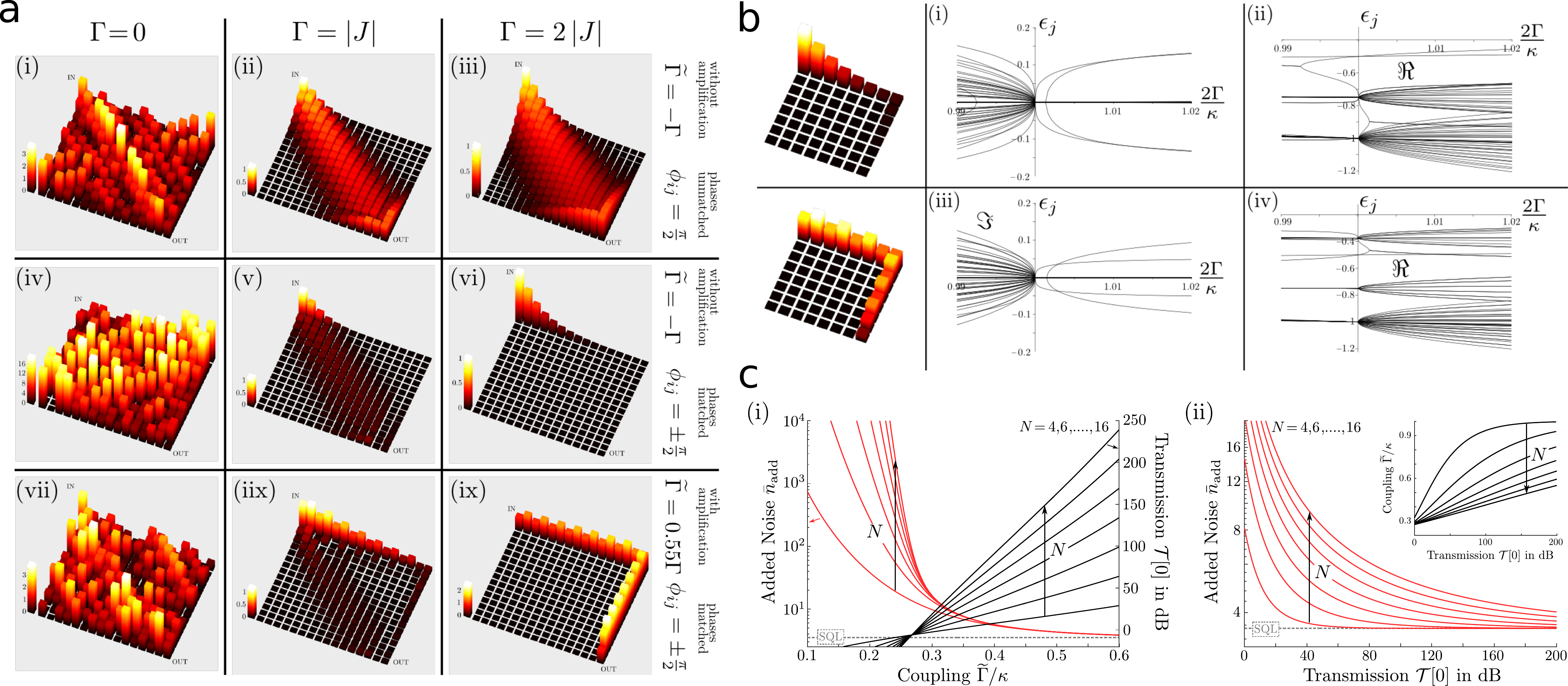} 
	\caption{Characteristics for nonreciprocal signal propagation along one edge.
	        \textbf{a} Averaged steady-state amplitude of each oscillator in a lattice of 256 cavities. The coherent hopping strength is fixed, i.e., $|J| = \kappa/4$ and $|J| = 0.55 \times \kappa/4$ for the amplification links if applicable, while the
	        dissipative coupling strength is increased from the left to right column. Once the directionality conditions are matched, i.e.,
	        graph (vi) and (ix), only the edge cavities have a finite occupation. However, to have a transmission close to
	        unity, amplification stages have to be implemented.
	        \textbf{b} Eigenvalues for a $N=8$ oscillator lattice without and with amplification.
	        The coherent hopping strength is set to $|J| = \kappa/4$ and the dissipative rate $\Gamma$ is varied.
	        The dynamics at the point of directionality is described by purely real eigenvalues. Note, this fact is independent of the chosen propagation path and the lattice size, for details see \cite{SI}.
	        \textbf{c} Transmission and added noise for propagation over one edge. By increasing $\widetilde \Gamma/\kappa$ the signal gets amplified, while the added noise is suppressed.
  	         The transmission diverges for $\widetilde \Gamma/\kappa \rightarrow 1$. 
  	         In Figure c(i) the suppression of the added noise is rather independent of the lattice size. The reason herefor is, that a larger lattice size involves more amplification stages, thus, a larger coupling strength $\widetilde \Gamma$ results in a higher gain value. Comparing the added noise for various $N$ and fixed transmission value we see, that a larger lattice size requires
  	         a larger amount of gain to come close to the quantum  limit, cf. graph c(ii).
	          }
	\label{fig:PropOneEdge}
 \end{figure*}
%
\section{The 2D system}\label{Sec.:2Dsystem}
%
The consideration of nonreciprocal transmission in a two-dimensional lattice introduces an additional aspect. There is a large degree of freedom in designing nonreciprocal transmission between two ports attached to such a lattice, posing in principle a difficult optimization problem.  This large optimization space also harbors a unique opportunity to route excitations nonreciprocally through a path that is dynamically reconfigurable.   
The optimization space we consider consists of the choice of link variables $\{ \Gamma_{ij}, \phi_{ij} \}$ for dissipative hopping. Additionally, amplification stages via the interaction \Eq{Eq.PAHam} will be inserted between select nodes $\{ \widetilde{\Gamma}_{ij}, \tilde{\phi}_{ij} \}$. We aim for flexibility in how we choose to propagate through the lattice while keeping a simple pattern for the oscillator couplings. 

We first analyze a configuration that allows the nonreciprocal routing of the excitation around one edge of the lattice, as shown in \Fig{fig:VariousPropagationWays}(a). An analytic solution can be obtained for a configuration with uniform dissipative coupling strength $\Gamma_{ij} = \Gamma$ on all links ("dissipative links") and choosing the pattern of phases $\phi_{ij} = \pm \pi/2$ as shown in Fig.~\ref{fig:SketchOneEdge} for the example of a lattice of 16 oscillators. In addition we insert an amplifying link at every second link along the designated edge of propagation (except at the corners) and denote the effective coupling strength of two neighbors at these stages as $\widetilde \Gamma$. The latter is a measure of how strongly we amplify the signal on its way through that link. Impedance matching and nonreciprocal propagation is ensured for the choice $\Gamma = 2|J| = \kappa/2$. For these conditions, forward transmission is given by
\begin{align}\label{Eq.:TransmissionOneWayEdge}
 \mathcal T_{1\rightarrow N^2} =& \frac{1}{4} \left[ \frac{2 \widetilde \Gamma \kappa }{\left( \kappa - \widetilde \Gamma \right)^2} \right]^{2N-4},
\end{align}
from  which we infer that stability requires $\widetilde \Gamma < \kappa$. \Fig{fig:PropOneEdge}c (i) depicts the transmission as a function of $\widetilde \Gamma$ for various lattice sizes $N$, showing that considerable signal amplification can be attained while staying away from the instability condition $\widetilde \Gamma = \kappa$. 
To transfer a signal successfully through the lattice, the implementation of the amplification stages is crucial.
Without the latter, the propagation is still nonreciprocal and over the edge but the signal amplitude decays due to induced local damping, cf. Fig.~\ref{fig:PropOneEdge}(a) graph (vi) vs (ix).

 \begin{figure*}[t] 
 \centering\includegraphics[width=0.9\textwidth]{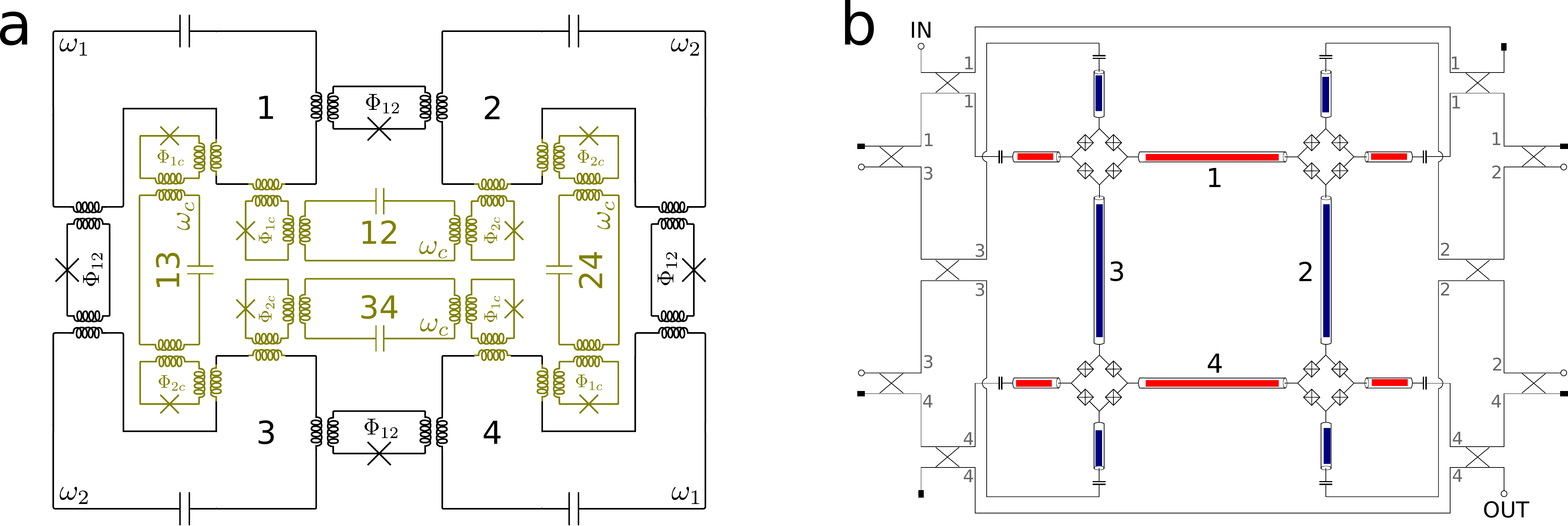} 
	\caption{Possible implementations for a $2\times 2$ lattice in sc-architectures.
	         \textbf{a} Circuit schematic for a lattice of LC-resonators. The oscillators are pairwise coupled via tunable couplers, i.e., a loop intersected by a Josephson junction.
	         Applying a flux $\Phi_{ij}$ to the loops realizes a tunable coupling $g_{ij}(t)$ between the oscillators. For resource efficiency we choose the  
	         link-oscillator (olive lines) to be degenerate in frequency ($\omega_c$), and the node-oscillators 1(2) and 4(3) are degenerate in frequency $\omega_{1}(\omega_{2})$ as well. 
	         Due to these degeneracy the 12 coupler loops require only three distinct flux values ($\Phi_{12}$, $\Phi_{1c}$ and $\Phi_{2c}$ as denoted in the graph).
	         \textbf{b} Example for an implementation utilizing Josephson parametric converters (JPC), a device which realizes three-wave mixing between the orthogonal modes
	         $X,Y$ and $Z$  of each JPC. For the lattice setup, the common mode $Z$ of each JPC corresponds to the link-oscillators, while the modes $X$ and $Y$ form the node oscillators (highlighted in red and blue). Note, in contrast to a standard JPC setup \cite{sliwa_reconfigurable_2015}, this design requires that each mircostrip resonator is intersected with two JPMs.
	         } 
	\label{fig:SketchImplementation}
 \end{figure*}

\section{Noise characteristics}\label{Sec:2DAmp}
%
Noise at the output port can be characterized by the symmetrized noise spectral density $\bar S_{N^2,\OUT}[\omega] = \frac{1}{2} \int \frac{d\Omega}{2\pi} \ev{\left\{\hat d_{N^2,\OUT}[\omega],\hat d_{N^2,\OUT}^{\dag}[\Omega]\right\} }$ which can be evaluated using input-output theory \cite{SI}.  
For example, taking the configuration discussed in the previous section (e.g. \Fig{fig:SketchOneEdge}), the added noise referred back to the input yields   
\begin{align}
	 \bar n_{\ADD} =& \frac{\bar S_{N^2} [0]}{ \mathcal T_{1\rightarrow N^2}} - \left(\bar n_{d_{1}}^T  + \frac{1}{2}\right)  \simeq  \  7 \left( \bar n_{\textup{link}}^T   +\frac{1}{2}\right) 
\end{align} 
for large gain, i.e.,  $\widetilde\Gamma/\kappa \rightarrow 1 $ and equal bath temperatures for all link-oscillators. 
Thus for zero temperature and large gain, the active lattice adds a minimum noise of 3.5 quanta to the signal. Inspection of \Fig{fig:PropOneEdge}c (i) reveals that this large-gain limit for noise has an asymptote that is independent of the size of the lattice $N$. 

There are three interesting aspects of the noise characteristics. Firstly, the latter strongly depend on the path of propagation. For instance, for propagation along one edge the minimum added noise as found above is 3.5 quanta, while for a design where nonreciprocal propagation takes place along {\it both} edges (\Fig{fig:VariousPropagationWays}(b)), the added noise is found to be 1.5 quanta \cite{SI}. This reduction in noise is due to destructive interference. From these two examples, it is clear that there is a rich optimization space for both maximizing the gain and minimizing the added noise. 

Another interesting aspect is that the noise characteristics at the output largely depend on the placement of the first amplification stages \cite{SI}. This is also the reason behind the observed insensitivity of the noise to the lattice size. On the other hand, the number of amplification stages directly contribute to the signal gain and increasing their number helps in keeping the operation point of individual amplification stages away from the instability point. We find that these are helpful rules of thumb in a future design of a reconfigurable lattice amplifier, but are far from exhausting the space of possibilities. 
%
%
\section{Implementation}\label{Sec.:Implementation}
%
In this section we discuss the physical implementation of the active lattice Hamiltonian \Eq{Eq.LatticeHam} with superconducting electrical circuits. 
We note that these effective interactions present in this Hamiltonians are linear. The implementations we discuss require the dynamic modulation of non-linear elements with multiple tones to generate the requisite linear interaction terms in an appropriate rotating frame. 

The first implementation we discuss is a lattice of LC-resonators which are pairwise connected via tunable coupler loops \cite{chen_qubit_2014,roushan_chiral_2017}, see Fig.~\ref{fig:SketchImplementation}a 
for the example of a circuit diagram of a $2\times2$ lattice. Each tunable coupling loop is intersected with a single Josephson junction and inductively coupled to two oscillators.  
Treating the loop by an external magnetic flux allows for a tunable coupling $g (t)$ between these oscillators.
This tunable coupling element between two oscillator modes $\hat a$ and $\hat b$ has been discussed \cite{chen_qubit_2014} and implemented experimentally \cite{roushan_chiral_2017}. 
In these setups a pair-wise interaction between the two modes can be generated, given by $\hH= g(t) (\hat a + \hat a^{\dag}) (\hat b + \hat b^{\dag})$. 
Harmonic modulation $g(t) = G \cos(\omega_p t + \phi)$ induces parametric processes between the modes. The choice of the pump frequency $\omega_p$ determines which 
processes are resonant, i.e., driving at the sum of the oscillators frequencies leads to amplification, while driving at the frequency difference induces frequency conversion.   
 
In the active lattice setup one unit cell consists of three oscillators which are nondegenerate in frequency.
The oscillators are pairwise connected via the tunable couplers realizing the lattice Hamiltonian in Eq.~\ref{Eq.LatticeHam} if all oscillator pairs are driven at the respective frequency difference of each pair. 
In principle, this would require three different pump frequencies $\omega_p$ per unit cell.
However, it is possible to reduce the number of pump sources by using second harmonics of the pumps \cite{kamal_minimal_2016} and by designing a dual-frequency node-oscillator lattice with
link-oscillators which are degenerate in frequency, see Fig.~\ref{fig:SketchImplementation}a. 

A second implementation is based on the Josephson parametric converter (JPC)\cite{abdo_nondegenerate_2013}, see Fig.~\ref{fig:SketchImplementation}b. 
The JPC realizes three-wave mixing and can be operated as a reciprocal or nonreciprocal quantum limited amplifier \cite{sliwa_reconfigurable_2015}.
The coupling elements of the JPC are based on the Josephson parametric converter (JRM), a ring which consist of four JJ-junctions arranged in a Wheatstone bridge configuration. Each JRM supports three orthogonal electric modes ($X,Y,Z$) and by treating the ring by a flux $\Phi$ close to half a flux quantum this device realizes three-wave mixing between these modes \cite{abdo_nondegenerate_2013}. Combining the JRM with microstrip resonators realizes the Josephson parametric converter (JPC), a purely dispersive device which realizes the quadratic interactions necessary for the active lattice setup.

The coupling between linear resonators could as well be realized via a superconducting interference device (SQUID). 
A SQUID-based tunable three-oscillator element has recently been used to implement a nonreciprocal frequency converter that can in-situ be reconfigured to a phase-preserving directional amplifier \cite{lecocq_nonreciprocal_2016}. 
A further design option involves the replacement of the link-oscillator by a strongly damped qubit or a mechanical resonator. 
The latter could be phononic modes of an optomechanical crystal \cite{fang_generalized_2017} or a electromechanical drum-head resonator \cite{bernier_nonreciprocal_2016}.  
Further details for the case of a JPC and a qubit implementation are found in the SI.  
%
%
\section{Conclusion}
Optimal routing of signals in a large-scale quantum information processor comes with certain requirements: on-chip implementation of as many components as possible, quantum limitedness, robustness to imperfections and protection of signal sources from unwanted back reflections. We introduced a 2D superconducting circuit architecture which allows the nonreciprocal routing of excitations at the quantum level, and which can achieve all the desired characteristics. We showed that by engineering the interactions in a dual-lattice design we obtain full control over the propagation path of a signal injected into the structure. 

An important advantage of the proposed active lattice architecture is that the use of parametric interactions takes the load off the required fabrication uniformity over individual components. The latter is often considered a serious problem in scaling up to larger architectures. However, the implementation of the proposed active lattice requires a layout that will allow the full control over the modulated elements, while keeping their cross-talk to a minimum. This is a formidable challenge that all quantum information processing schemes have to face. A promising route in this direction is a multilayer architecture which would place the lattice on one chip and the control lines for modulation in another layer. Efforts in this direction are underway in several laboratories \cite{brecht_multilayer_2016, minev_planar_2016, roushan_chiral_2017}. 
 
%

%
%
\clearpage

\global\long\def\theequation{S.\arabic{equation}}

\global\long\def\thefigure{S.\arabic{figure}}

\setcounter{equation}{0}

\setcounter{figure}{0}

\thispagestyle{empty}
\onecolumngrid
\begin{widetext}
\begin{center}
{\fontsize{12}{12}\selectfont
\textbf{Supplemental Material for \\[1mm]
``Nonreciprocal Signal Routing in an Active Quantum Network"\\
[5mm]}}
{\normalsize A. Metelmann and H. E. T\"ureci\\[1mm]} 
{\fontsize{9}{9}\selectfont  
\textit{Department of Electrical Engineering, Princeton University, Princeton, New Jersey 08544, USA}} 
 \end{center}   \normalsize  
%
%
%
\section*{One-way transport between two connected nodes}
%
We first discuss the basic ingredients of the proposed active lattice, namely the implementation of one-way transport between two given nodes $i \rightarrow j$.  We start out from the effective lattice Hamiltonian given in Eq.(1) of the main text and derive the Heisenberg-Langevin equations of motion for two sites $i$ and $i$ and the link oscillator $ij$:
\begin{align}
\dot{\hat{d}}_i = & - i G_{ij} \ex{-i \phi_{ij}} \dj - i G_{i;ij} \dij,
\nonumber \\
\dot{\hat{d}}_j  =&  - i G_{ij} \ex{i \phi_{ij}} \di - i G_{j;ij} \dij,
 \nonumber \\
\dot{\hat{d}}_{ij}  =& - \frac{\kij}{2} \dij - \sqrt{\kij} \hat d_{ij,\IN} - i G_{i;ij} \di - i G_{j;ij} \dj,
 \end{align}
here we assumed that the link-oscillator is coupled to an equilibrium bath with rate $\kij$,
and that $\kij$ is the dominant loss channel in the system, comparable to $G_{ij}$; $\hat d_{ij,\IN}$ describes thermal and vacuum noise driving the link-oscillator. As discussed below, the satisfaction of this condition is no requirement for directionality, but will suppress the reverse propagating signal for a broader range of frequencies. Adiabatically eliminating $\dij$, we obtain the Heisenberg-Langevin equations of motion for the two sites $i$ and $j$ 
\begin{align}\label{Eq.EoMCouplingGeneralSI}
\dot{\hat{d}}_i = & - \frac{\Gamma_{i;ij}}{2}  \di + i \sqrt{\Gamma_{i;ij}}  \hat d_{ij,\IN} -  \left[ i G_{ij}  \ex{-i\phi_{ij} }  + \frac{\sqrt{ \Gamma_{i;ij} \Gamma_{j;ij} }}{2}  \right] \dj ,
\nonumber \\
\dot{\hat{d}}_j  =&  - \frac{\Gamma_{j;ij}}{2} \di + i \sqrt{\Gamma_{j;ij}}  \hat d_{ij,\IN} -  \left[ i  G_{ij} \ex{+i\phi_{ij} }  + \frac{\sqrt{ \Gamma_{i;ij} \Gamma_{j;ij} }}{2}  \right] \di ,
 \end{align} 
with the definitions $\Gamma_{n;ij} = 4 G_{n;ij}^2/ \kij, (n\in i,j)$. In this damped link-oscillator regime  the system of two node-oscillators can as well be described via a Markovian master equation, where the dissipative interaction is described via the non-local superoperator
\begin{align}
\Gamma_{i;ij}  \mathcal L \left[\hat d_{i} +  \sqrt{\frac{\Gamma_{j;ij}}{\Gamma_{i;ij}}} \hat d_{j}\right] \hat \rho,
\hspace{0.5cm}
\mathcal L [\hat o] \hat \rho = \hat o  \hat \rho \hat o^{\dag} - \frac{1}{2} \hat o^{\dag} \hat o \hat \rho - \frac{1}{2} \hat \rho \hat o^{\dag} \hat o.
\end{align}
Thus, the link-oscillator can be interpreted as an engineered reservoir for the node oscillators, which has two important effects. It gives rise to an indirect exchange term between two node oscillators, while inducing local damping at a rate $\Gamma_{n;ij}/2$.

We aim for the situation where the oscillator $j$ is driven by the oscillator $i$ but not vice versa. This can be achieved through balancing the effective dissipative hopping term, 
i.e., the second term in the square brackets in Eq.~(\ref{Eq.EoMCouplingGeneralSI}), with the unitary hopping term  \cite{metelmann_nonreciprocal_2015_SI}.  
The balancing conditions become 
\begin{align}
 \phi_{ij} \equiv  -\frac{\pi}{2} \quad \text{and} \quad    G_{ij} \equiv \frac{\sqrt{ \Gamma_{i;ij} \Gamma_{j;ij} }}{2}.
\end{align}
These conditions provide a manifestly directional coupling for excitations ($j$ is coupled to $i$ but not vice versa):  
\begin{align}\label{Eq.TwoModeDecoupled}
\dot{\hat{d}}_i = &  - \frac{\Gamma_{i;ij}}{2}  \di +  \sqrt{\Gamma_{i;ij}}  \hat d_{ij,\IN}  ,
\nonumber \\
\dot{\hat{d}}_j  =&  - \frac{\Gamma_{j;ij}}{2}  \di +  \sqrt{\Gamma_{j;ij}}  \hat d_{ij,\IN} -  \sqrt{ \Gamma_{i;ij} \Gamma_{j;ij} }   \di .
 \end{align}
This mechanism should be contrasted to earlier work in lattices subject to artificial gauge fields \cite{roushan_chiral_2017_SI}, where dissipation is desired to be minimal. In the latter case, directional propagation is a pure interference effect, can be described through Hamiltonian dynamics and therefore it is more appropriate to talk about the breaking of time-reversal symmetry. 
%
\section*{Nonreciprocal propagation along a 1D chain}
%
In this section we provide further details for the example of nonreciprocal transport along a chain of oscillators. The oscillator chain consist of $N$ oscillators with alternating resonant frequencies $\omega_{1,2}$. Each oscillator pair is coupled via a coherent hopping interaction and we assume equal couplings $G_{ij} e^{-i\phi_{ij}} \equiv J$ between oscillator $i$ and $j=i+1$. 
For better overview, we as well assume a uniform coupling to the link-oscillators, i.e., we set $ G_{i;ij} \equiv \lambda $; and we denote the link-oscillators here as $\hat d_{i i+1} \equiv \hat c_{i}$.
Moving to a frame with respect to each oscillator's resonant frequency the final Hamiltonian reads
\begin{align}\label{Eq.:SIchainHam}
\hH = \sum_{i = 1}^N \delta_i \hat d_{i}^{\dag} \hat d_{i} + \left[ J \sum_{i}^{N-1} \hat d_{i}^{\dag} \hat d_{i+1} 
       + \lambda \sum_{i}^{N-1} \hat c_{i}^{\dag}   \left(\hat d_{i}  + \hat d_{i+1}\right) + h.c \right],
\end{align}
here we take into account the detunings $\delta_{i} = \omega_i - \omega_{1,2}$. These detunings can be a consequence of the oscillators' resonant frequencies deviating from $\omega_{1,2}$ or from
those of the external driving frequencies that are necessary to obtain the interactions in Eq.(\ref{Eq.:SIchainHam}) as further discussed in Sec.~\ref{Sec.:SIimplementation}.

We assumed that oscillator $1$ and  $N$ are coupled to external waveguides with coupling strength $\kappa_{e}$, additionally, all oscillator are coupled to Markovian baths with coupling strength $\eta$. The latter is introduced to evaluate the impact of finite loss acting on the node oscillators. We adiabatically eliminate the link-oscillators and apply the directionality condition $J  = i \frac{\Gamma}{2}$ with $\Gamma = 4 \lambda^2/\kij$.
By impedance matching the system via $\Gamma  = \kappa$ we obtain the equation system 
\begin{align} 
	 \frac{d}{dt} \hat d_1  =& - \sqrt{\kappa_{e}} \hat d_{1,\IN} 
	                           - \sqrt{\eta} \hat \xi_{1,\IN}  + i  \sqrt{\kappa}  \hat c_{1,\IN} 
	                           - \left( i \delta_1 +  \kappa   \right) \hat d_1   ,
	\nonumber \\ 
 	\frac{d}{dt}  \hat d_m  =& - \sqrt{\eta} \hat \xi_{m,\IN}  + i \sqrt{\kappa} \hat c_{m-1,\IN} + i \sqrt{\kappa} \hat c_{m,\IN} 
 	                          - \left(  i \delta_m    + \frac{\eta + 2\kappa }{2} \right) \hat d_m 
 	                          -  \kappa  \hat d_{m-1} , \hspace{0.2cm} m \in [2, N-1]
         \nonumber \\
        \frac{d}{dt}  \hat d_N  =&  -  \sqrt{\kappa_{e}} \hat d_{N,\IN} - \sqrt{\eta} \hat \xi_{N,\IN} + i \sqrt{\kappa} \hat c_{N-1,\IN}
                                      - \left( i \delta_N  + \kappa\right) \hat d_N  
				         -   \kappa \hat d_{N-1}    
\end{align}
with $\kappa = \kappa_e + \eta$. $\hat \xi_{i,\IN}$ describe thermal and vacuum fluctuation impinging on each oscillator, while $\hat c_{m,\IN}$ correspond to the noise contribution arising due to the coupling to the link-oscillators.
Although, we obtain a system of $N$ coupled equations, it is possible to obtain analytic expressions for the scattering parameters.
Using input-output theory, $\hat d_{i,\OUT} = \hat d_{i,\IN} + \sqrt{\kappa_e} \hat d_{i}$, we can derive the transmission coefficient
\begin{align}
 \mathcal T[0] =   \frac{\left(1 - \frac{\eta}{\kappa}\right)^2  }{\left[ 1 +   \frac{\delta_1^2}{\kappa^2}  \right] \left[1 +   \frac{\delta_N^2}{\kappa^2}   \right] } 
	   \prod_{m =2}^{N-1} \frac{1}{\left[ 1 + \frac{\eta}{2\kappa} \right]^2  +   \frac{\delta_m^2}{\kappa^2} }
	   \overset{\delta_i \equiv \delta }{=}
	     \left[ \frac{ 1 - \frac{\eta}{\kappa}  }{  1 +   \frac{\delta^2}{\kappa^2}     } \right]^2 
	      \left[ \left( 1 + \frac{\eta}{2\kappa} \right)^2  +   \frac{\delta^2}{\kappa^2}  \right]^{2-N}
	      \simeq 1 - N \left(\frac{\eta}{\kappa} + \frac{\delta^2}{\kappa^2}  \right),
\end{align}
with $\hat d_{N,\OUT} =  t[0] \, \hat d_{1,\OUT}$ and $ \mathcal T[0] \equiv \left| t[0] \right|^2$. To reach close to unity transmission, the detunings $\delta/\kappa$  and the intrinsic losses $\eta/\kappa$ have to be kept at minimum. The length of the oscillator chain becomes important, as with the number of oscillators exposed to decay channels the number of loss channels increases. On the other hand, having finite detunings and intrinsic losses does not impact the nonreciprocity of the system.  
This can be seen by considering the output of the first oscillator
\begin{align}
 	   \hat d_{1,\OUT}  =&      \frac{ \frac{\eta}{\kappa} + i \frac{\delta_1}{\kappa}   }{1 + i \frac{\delta_1}{\kappa} } \hat d_{1,\IN} 
 	                           - \frac{\sqrt{1 - \frac{\eta}{\kappa} }}{1 + i \frac{\delta_1}{\kappa} }\left[   
	                             \sqrt{ \frac{\eta}{\kappa}  }   \hat \xi_{1,\IN}  - i     \hat c_{1,\IN} \right] ,
\end{align}
crucially, the output does not contain any contributions from oscillators higher up of the chain. The first term simply denotes the reflection of an input signal $\hat d_{1,\IN} $, while the second term describes the noise contribution from the first link-oscillator and the bath of oscillator 1. However, one still aims for a small output of oscillator 1 to protect the source providing the input signal, if that is desired.
For the optimal case of $\eta/\kappa, \delta_1/\kappa \rightarrow 0$ the output simply becomes $\hat d_{1,\OUT} = i \hat c_{1,\IN}$, i.e., even in a perfect setting we have an effective noise temperature at the output oscillator 1 which is determined by the first engineered reservoir. Hence, a cold bath driving the link-oscillator is desirable.

A further crucial  aspect is the noise which is added to the signal while passing down the oscillator chain. The total output of the oscillator $N$ is given by 
\begin{align}
 \hat d_{N,\OUT}  =&   
	        \frac{\eta}{\kappa}   \hat d_{N,\IN}    
              - \sqrt{\frac{ \eta  }{\kappa} - \frac{\eta^2}{\kappa^2} }     
              \left\{\hat \xi_{N,\IN}  +  \sum_{k = 2}^{N-1}  
  	    \frac{(-1)^{N-k}  }{\left[1   + \frac{\eta}{2\kappa}\right]^{N-k}}  
  	    \left[  \hat \xi_{k,\IN} + \frac{i}{2} \sqrt{\frac{\eta}{\kappa}}  \hat c_{k,\IN} \right] 
  	    +    \frac{ (-1)^{N-1} }{\left[1 + \frac{\eta}{2\kappa}\right]^{N-2}       } 
  	          \hat \xi_{1,\IN}\right\} 
	   +  t[0] \hat d_{1,\IN}
\end{align}
here we considered the limit $\delta_{i}/\kappa \rightarrow 0$ as the expression including finite detunings is rather cumbersome.
The output contains contributions from all oscillators, except for the first link-oscillator as this contribution ends up in the output of oscillator 1, as discussed above.
To characterize the noise properties we calculate the symmetric noise spectral density, defined as
\begin{align}
 \bar S_{N,\OUT}[\omega] = \frac{1}{2} \int \frac{d\Omega}{2\pi} \ev{\left\{\hat d_{N,\OUT}[\omega],\hat d_{N,\OUT}^{\dag}[\Omega]\right\} },
\end{align}
for the evaluation we use the noise correlators
$\ev{\hat o_{\rm in}^{\PD}(\omega)\hat o_{\rm in}^{\dag}(\Omega)}= \ev{\hat o_{\rm in}^{\dag}(\omega) \hat o_{\rm in}^{\PD}(\Omega)} + 2 \pi \delta(\omega + \Omega)  = 2 \pi \delta(\omega + \Omega) (\bar n_o^T + 1)$, where $o = \xi_{i}, c_{k},d_{1,N}$.
The output noise spectral density on resonance yields
\begin{align}
  \bar S_{N,\OUT}[0]  =&  \frac{1}{2}  
	      +  \frac{ \left[  1 - \frac{\eta}{\kappa}  \right] }{\left[ 1 + \frac{\eta}{2\kappa} \right]^{2(N-2)}}
	        \left\{ \left[  1 - \frac{\eta}{\kappa}  \right] \bar n_{d_{1}}^T   +  \frac{ \eta  }{\kappa}        \bar n_{\xi_{1}}^T    \right\} 
	      +   \frac{\eta }{\kappa } \left[
	         \frac{\eta}{\kappa }  \bar n_{d_{N}}^T   +  \left[1 - \frac{\eta}{\kappa} \right]   \bar n_{\xi_{N}}^T    
              +   \left[1 - \frac{\eta}{\kappa} \right]   
                \sum_{k = 2}^{N-1}  
  	    \frac{\left[  \bar n_{\xi_{k}}^T      +   \frac{\eta}{4\kappa}  \bar n_{c_{k}}^T    \right] }{\left[1   + \frac{\eta}{2\kappa}\right]^{2(N-k)}}   \right], 
\end{align}
here the second term describes the thermal noise originating from oscillator 1; while the terms in the square brackets denote noise contributions from the remaining node oscillators and link-oscillators.
For the case of negligible intrinsic losses, the contribution which is always present is thermal noise associated with the input signal, i.e., $\bar S_{N,\OUT}[0] \rightarrow 1/2 + \bar n_{d_{1}}^T$ for $\eta/\kappa \rightarrow 0$. However, if one assumes thermal baths $\bar n^T$ with equal temperatures for all oscillators, the output spectrum is independent of the ratio $\eta/\kappa$, $\bar S_{N,\OUT}[0]  =   \left(\bar n^T   + \frac{1}{2}\right)$. Crucially, the thermal noise contribution of the link-oscillators scales quadratically with $\eta/\kappa$. This becomes clearer if we set $ \bar n_{d_{1,N},\xi_{i}}^T   \equiv  \bar n_{d}^T  $ and $ \bar n_{c_{k}}^T \equiv \bar n_{\rm link}^T $ and expand the output noise for small values of $\eta/\kappa$ 
\begin{align}
  \bar S_{N,\OUT}[0]  
               =& \frac{1}{2} + \bar n_{d}^T     
                     +  \frac{1}{2}   \left[    \frac{N}{2}  - 1      \right] \left(   \bar n_{\rm link}^T -  \bar n_{d}^T \right)  \frac{\eta^2}{\kappa^2}  
                     + \mathcal O \left[  \frac{\eta^3}{\kappa^3} \right] .
\end{align}
The reason for this quadratic scaling lies in an interference effect; neighboring node-oscillators are coupled to the same link-oscillator and hence, part of the noise originating from the link-oscillator cancels out.
 
%
\section*{2D system: example of a 16 oscillator lattice}
%
%
On the basis of a 16 oscillator lattice we discuss in this section the details of nonreciprocal signal propagation in two dimensions. By tuning the coupling strengths and the phases in our setup we can choose an arbitrary path through the structure. We focus on three different paths as depicted in Fig.~\ref{fig:Sketch16Cavities}. The arrows between each oscillator denote the direction of the signal/information transfer, which is determined via the phase $\varphi_{ij}$. Depending on the chosen direction the latter take the following values  
\begin{align}
 \downarrow, \rightarrow : \varphi_{ij} = - \frac{\pi}{2},
 \hspace{0.5cm}
 \uparrow,  \leftarrow  :  \varphi_{ij} = + \frac{\pi}{2}.
\end{align}
We assume that each oscillator pair couples with the same strength to their respective link-oscillator,  i.e., we set $\Gamma_{n,ij} \equiv \Gamma_{ij}$.
Hence, the remaining directionality condition between two oscillators simplifies to $G_{ij} = \Gamma_{ij}/2$. 
After applying the latter condition we end up with an effective uni-directional coupling of oscillator $i$ and $j$ with strength $\Gamma_{ij}$, cf. Eq.~(\ref{Eq.TwoModeDecoupled}).
The respective values for these effective couplings  $\Gamma_{ij}$ are denoted in Fig.~\ref{fig:Sketch16Cavities}.
 \begin{figure}  
 \centering\includegraphics[width=1.0\textwidth]{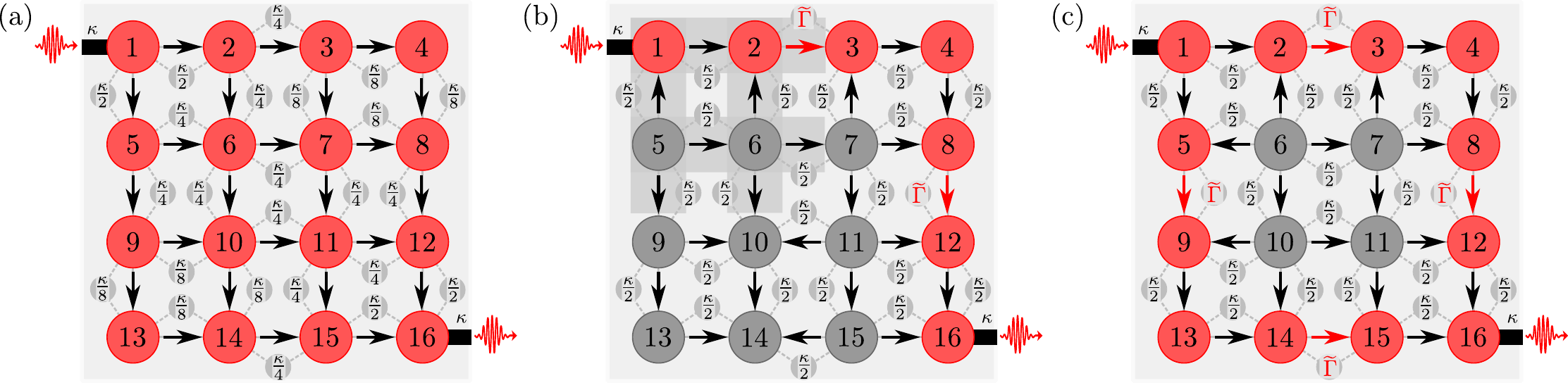} 
	\caption{Sketch of the 16 oscillator lattice. Three different ways to propagate through the structure are highlighted in yellow.
	         The black arrow indicates which direction is chosen via matching the phase $\phi_{ij}$ and the effective strength of the interactions $\Gamma_{ij}$. 
	         The respective values for the latter are denoted in the graph, where $\Gamma_{ij} = \widetilde \Gamma$ corresponds to an amplification stage between two modes.
	         The darker gray background in graph (c) denotes the oscillators and reservoirs who contribute to the added noise.}
	\label{fig:Sketch16Cavities}
 \end{figure}
\subsection*{Hopping over all oscillators}

We start with the illustration of the special case of hopping over all oscillator nodes, as sketched in Fig.~\ref{fig:Sketch16Cavities}(a).
Here the signal passes through all oscillators, and all phases are set to $\varphi_{ij} = - \frac{\pi}{2}$, i.e., the signal propagates 
to the right- and downwards. Although, propagation over all oscillators is not a distinct propagation path, it comprised a couple of interesting properties worth pointing out. As mentioned in the main text, the point of directionality here marks a transition from oscillatory to purely damped dynamics (in an appropriately rotated frame). Moreover, it allows for unity transmission without involving an amplification stage, i.e., it is possible to match the sum of the local damping experienced by each mode to the coupling to its neighbors from which it receives the input signal. This means that a signal has to be passed on faster than it can leak out of an oscillator.  
 
To illustrate the resulting pattern of coupling strengths further, we consider the signal oscillator $1$ and its right neighbor oscillator $2$, cf. Fig.~\ref{fig:Sketch16Cavities}(a).
Assuming that all directionality conditions are matched their expectation values evolve as
\begin{align}
 	 \frac{d}{dt}  \bar d_1  =& -  \frac{\kappa + \Gamma_{12}+ \Gamma_{15}}{2}  \bar d_1  - \sqrt{\kappa} d_{1,\IN} ,
	\nonumber \\ 
 	\frac{d}{dt}   \bar d_2  =&  - \frac{  \Gamma_{12} + \Gamma_{23} + \Gamma_{26}}{2}  \bar d_2  
 	                               -    \Gamma_{12}  \bar d_1    ,
\end{align}
here the coupling to the external waveguide of oscillator $1$ is associated with local damping in the amount of $\kappa/2$. Additionally, 
the dissipative couplings to oscillator $2$ and $5$ result in  local dampings $\Gamma_{12,15}/2$ respectively. 
For impedance matching we need $\Gamma_{12} + \Gamma_{15} \equiv \kappa$ to kill the reflection of the input signal. A symmetric choice is simply $\Gamma_{12} = \Gamma_{15} = \kappa/2$.
The latter fixes part of the local damping the neighboring oscillators experience. On the other hand, oscillator $2$'s local damping equals $(\Gamma_{12} + \Gamma_{23} + \Gamma_{26})/2$, while its coupling to oscillator $1$ is $\Gamma_{12} = \kappa/2$. To transfer the input signal fast enough we
need $\Gamma_{23} + \Gamma_{26} = \kappa/2$, leading to the stationary solution $\bar d_{2} = - \bar d_{1}$, which is exactly what we were after. It is important to note, that the effective coupling rate between two oscillators has decreased, i.e., for symmetric choice
we have $\Gamma_{23} = \Gamma_{26} = \kappa/4$. 
Overall, to match the local damping and the effective coupling between all oscillators for the whole 16 cavities setup, leaves us
with a pattern of staggered coupling strengths as illustrated in Fig.~\ref{fig:Sketch16Cavities}(a). This results in perfect transmission of the input signal, i.e., we have $\bar d_{16,\OUT} = - d_{1,\IN}$
as desired. This kind of pattern works as well for larger lattices, with the simple rule that along the upper edge the 
effective coupling between two neighbors $j$ and $j+1$ decreases as $\kappa/2^{j}$, reaching $\kappa/2^{N-1}$ at the corner. 
The same holds for the left edge and is reversed for the lower and right edge, while the inner couplings have to be adjusted accordingly. 

The discussed pattern for the coupling strengths results in unity transmission, as denoted above. 
However, under realistic conditions each oscillator would experiences losses due to the coupling to its environment. To study the influence of intrinsic losses, we couple each oscillator to a Markovian bath with rate $\eta$. 
The transmission becomes
\begin{align}
 \mathcal T_{\rm hopp}[0] =&   \left( \frac{   \kappa - \eta   } {4\eta + 3  \kappa   }   \frac{  \kappa^4}{\left[\eta + \kappa\right]^4}
	                              \left[ 1+  \frac{\eta + \kappa} { 2 \eta + \kappa }   
	                                +   \frac{1}{4}   \frac{\left[\eta + \kappa\right]^2}{\left[2\eta +  \kappa\right]^2 } \left( 1   +      \frac{4\eta + 3  \kappa}{4 \eta +  \kappa  }  \right)
	                               \right] \right)^2
                      = \; 1 - 16 \frac{  \eta }{\kappa } + \frac{428 }{3  } \frac{  \eta^2}{ \kappa ^2}+ \mathcal O\left[ \frac{\eta^3}{\kappa^3} \right]      ,         
\end{align}
in the second step we expanded the result for small ratios $\eta/\kappa$.
In the limit  $\eta/\kappa \rightarrow 0$ we have unity transmission. The expanded expression of the transmission for finite losses makes clear what happens
when the signal passes through the lattice: we have $16$ intermediate oscillators, thus in every of these oscillators we loose approximatively $\eta/\kappa$-part of the signal. 
The corresponding added noise follows the same logic 
\begin{align}
 \bar n_{\rm add} =   16 \left(   \bar n_{d}^T + \frac{1}{2} \right)    \frac{\eta  }{\kappa } 
                    + \left\{  \frac{ 293 }{3}  \left(  \bar n_{d}^T + \frac{1}{2} \right)
                             + \frac{ 563 }{36  }  \left( \bar n_{\rm link}^T + \frac{1}{2} \right)  \right\} \frac{\eta^2 }{\kappa^2 } 
                    + \mathcal O\left[ \frac{\eta^3}{\kappa^3} \right]   ,
\end{align}
where we assumed equal thermal baths $\bar n_{\rm link}^T$ for all link-oscillators, as well as for all node-oscillators $(\bar n_{d}^T)$.
Every node-oscillator contributes at least their vacuum fluctuations to the added noise. The contribution from the fluctuations of the link-oscillators is less damaging as it scales with $(\eta/\kappa)^2$. However, for $\eta/\kappa = 0$ we have no added noise corresponding to the optimal situation.

While the intrinsic losses do influence the transmission, they do not affect the directionality of the system.
This can easily be seen, if we calculate the spectral output noise density for oscillator 1, on resonance we obtain
\begin{align}
  \bar S_{1,\OUT}[0]     
                 =&  \frac{1}{2} + \bar n_{\rm link}^T  
                       +  \frac{\eta}{\kappa}   \left(  \bar n_{\xi_1 }^T - \bar n_{\rm link}^T \right)
                       +  \frac{ \eta^2  }{\kappa^2}     \left(  \bar n_{d_1}^T  - \bar n_{\xi_1}^T \right)         ,                   
\end{align}
where $\bar n_{\xi_1}^T$ denotes the averaged thermal occupation of the intrinsic oscillator-1 bath and
$\bar n_{d_1}^T$ is  associated with the thermal fluctuations accompanying a possible input signal.  
The output noise of oscillator 1 contains contributions from the link-oscillators which couple it to oscillator 2 and 5.
Crucially, besides the two reservoirs and the fluctuations impinging on oscillator 1, there is no further contribution
from other node-oscillators, i.e., oscillator 1 is perfectly decoupled from the lattice.
\subsection*{Propagation along one edge}
Next we consider propagation along one edge as sketched in Fig.~\ref{fig:Sketch16Cavities}(b). 
Here we can avoid a staggered coupling pattern and set all effective couplings to $\kappa/2$ (this directly ensures impedance matching), except at the amplification stages which we plant
between oscillators 2 and 3 as well at oscillators 8 and 12. For the latter case we choose $\Gamma_{ij} = \widetilde \Gamma$ for the effective couplings. We have to set the phases $\phi_{ij}$ in the right manner to ensure propagation only along  the edge. Consider for example the signal input oscillator, i.e.,  oscillator 1, we want to transmit the whole signal to oscillator 2, thus we
choose the phase $\phi_{12}$ in the way that oscillator 1 is decoupled from oscillator 2, while oscillator 2 is driven by oscillator 1, i.e., we set $\phi_{12} = -\pi/2$. 
To avoid that the any information from oscillator 1 is transmitted
to oscillator 5 we have to decouple oscillator 5 from oscillator 1 and thus set $\phi_{15} = +\pi/2$, i.e., the signal cannot enter oscillator 5. However, this comes with the price that oscillator 1 is driven by the noise impinging on oscillator 5. Clearly, we cannot implement an amplification step here, as it would lead not only to reflection of an input signal, but as well to amplification of the reflected signal. This is a situation we want to avoid.

The phases and coupling strengths are set as denoted in Fig.~\ref{fig:Sketch16Cavities}(b). Crucially, one has to fix all couplings between oscillators surrounding the propagation path.
However, the remaining couplings and phases are less crucial, e.g., for the example of the 16 oscillator lattice the lower left block of 4 oscillators (9,10,13,14) can be completely decoupled
and the couplings between these oscillators can be set arbitrarily. 
With this coupling scheme the transmission coefficient from the output of oscillator 16 (for $\eta = 0$)
\begin{align} \label{Eq.:Transmission16Cavities}
 \mathcal T_{\rm 1 edge}[0] = \frac{1}{4} \left[ \frac{2 \widetilde \Gamma \kappa}{\big( \kappa - \widetilde \Gamma\big)^2}\right]^4, 
\end{align} 
from  which we see that stability requires $\widetilde \Gamma < \kappa$. 
Without the amplification processes we would loose most of the signal, i.e., corresponding to setting $\widetilde \Gamma \rightarrow - \widetilde \Gamma$ in Eq.(\ref{Eq.:Transmission16Cavities}).
We do not need to be close to the instability at $\widetilde \Gamma = \kappa $ to transmit the signal with
large gain, e.g., setting $\widetilde \Gamma = \kappa/2$ we already have around $18$~dB of gain. 
However, including intrinsic losses results in a reduced gain, but the transmission is more robust in this case, as the propagating signal is much larger due to the amplification steps.

The remaining question is how much noise is added to the signal. As discussed in the main text, the noise contribution added to the signal before an amplification stage is entered sets the minimum of the added noise. 
This is simple to understand: if the added noise is already larger than the noise added before the gain stage we can never reach the quantum limit, as
the noise added before the amplification gets amplified in the same manner  as the signal and cannot be suppressed in any way. 
For the case of propagation over one edge the first amplification stage is between oscillators 2 and 3. 
Thus we have to consider the stationary solution for the EoM of oscillator 2  
\begin{align}
	  \hat d_2  =&      \;  \sqrt{ \mathcal G_2}   
	 	                     \bigg\{    \hat d_{1,\IN}  + i  2 \sqrt{\frac{\widetilde \Gamma}{\kappa}  }   \hat d_{23,\IN}^{\dag}  
	                             - i     \sqrt{\frac{1}{2}} 
	                             \left[    \hat d_{67,\IN}   +   \hat d_{610,\IN}
	                                   -   \hat d_{12,\IN}  -   \hat d_{26,\IN}
	                                   - \frac{1}{3} \left(   \hat d_{15,\IN}  + \hat d_{56,\IN} \right)
	                                   - \frac{4}{3}  \hat d_{59,\IN}\right]  \bigg\},	                                   
\end{align}
the first terms in the wavy brackets is simply the input signal, while all remaining terms denote noise contributions from the surrounding link-oscillators (we assumed $\eta = 0$),
see Fig.~\ref{fig:Sketch16Cavities}(b) where the gray highlighted area denotes all link-oscillators which contribute to the noise. 
The pre-factor of the upper expression is simply an intermediate gain factor which we labeled as $\sqrt{\mathcal G_{2}} = \sqrt{\kappa}/ (\kappa-\widetilde \Gamma)$.
The symmetrized noise spectra for oscillator 2 becomes on resonance 
\begin{align}
	  \bar S_2 [0] =&      \;    \mathcal G_2  \left(\bar n_{d_{1}}^T  + \frac{1}{2}\right)
	                         +  \mathcal G_2
	 	                       \bigg\{  \frac{4\widetilde\Gamma}{\kappa}  +    3  \bigg\}
	                                   \left(  \bar n_{\rm link}^T + \frac{1}{2} \right),
\end{align}
where we assumed equal bath temperatures for all link-oscillators.  
We can now extract the noise which is added up to the input signal at this stage; the added noise referred back to the input yields
\begin{align}
 \bar n_{2,\rm add} = \frac{\bar S_2 [0]}{\mathcal G_2} - \left(\bar n_{d_{1}}^T  + \frac{1}{2}\right) 
                   = \bigg\{  \frac{4\widetilde\Gamma}{\kappa}  +    3  \bigg\} \left(  \bar n_{\rm link}^T + \frac{1}{2} \right)
                   \rightarrow 7 \left(  \bar n_{\rm link}^T + \frac{1}{2} \right),
\end{align} 
in the second step we assumed the large gain limit, i.e., $\widetilde\Gamma \rightarrow \kappa$. 
The latter coincides with the final minimum of added noise, i.e., the overall lower limit for the noise added to the signal leaving the output of oscillator 16.
Thus, for zero temperature bath and large gain, we have a minimum noise of $3.5$ quanta added to the signal.      
The later limit is independent of the lattice size as long the first amplification stage is implemented between oscillator 2 and 3.
%
%
 \begin{figure}  
 \centering\includegraphics[width=1.0\textwidth]{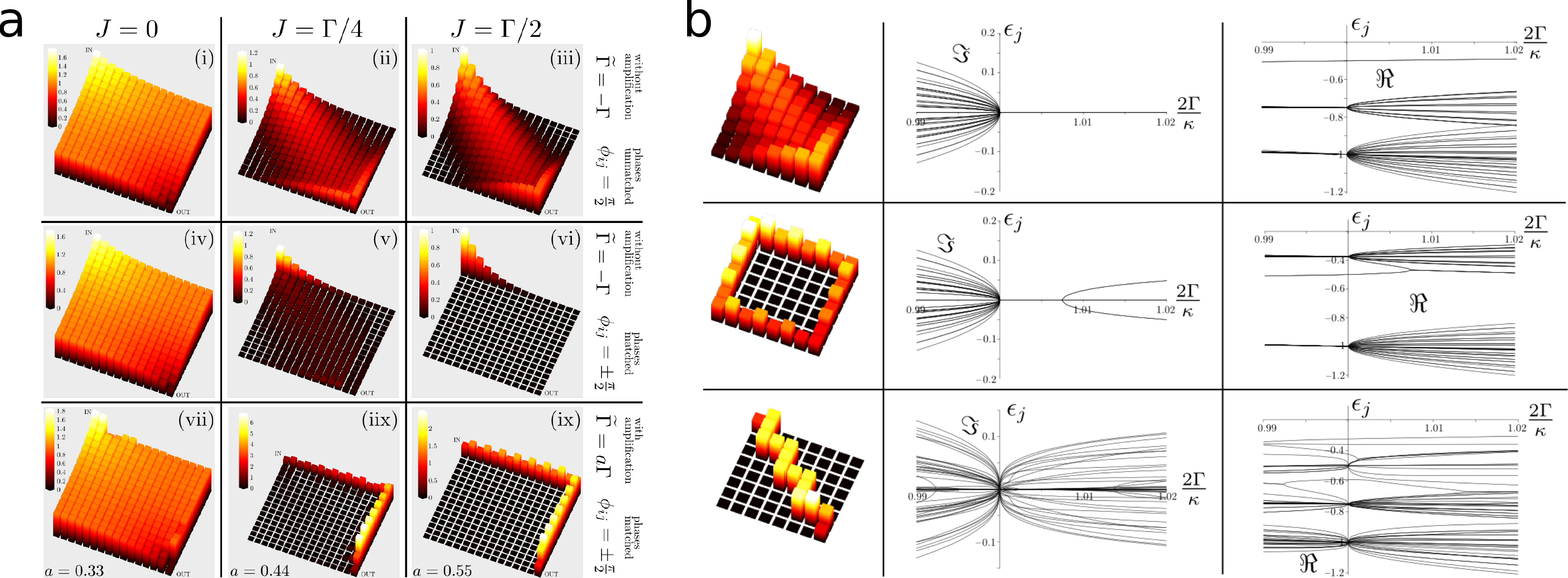} 
	\caption{\textbf{a} Averaged steady-state amplitude of each oscillator in a lattice of 256 cavities. The dissipative hopping strength $\Gamma$ is fixed, while the
	        coherent coupling strength $J$ is increased from the left to right column. Once the directionality conditions are matched, i.e.,
	        graph (vi) and (ix), only the edge cavities have a finite occupation. However, to have a transmission close to
	        unity, amplification stages have to be implemented.
	        \textbf{b} Eigenvalues as a function of dissipative coupling strength for a $N=8$ oscillator lattice for various propagation ways. 
	        All eigenvalues are real at the point of directionality $\Gamma = \kappa/2$.  
	        The coherent hopping strength is set to $|J| = \kappa/4$ and the dissipative rate $\Gamma$ is varied.
	        The dynamics at the point of directionality is described by purely real eigenvalues, the latter is independent of the chosen propagation path and the lattice size. 
  	         }
	\label{fig:VarPaths}
 \end{figure}
%
\subsection*{Propagation along both edges}
As a final example for nonreciprocal signal propagation in a 2D lattice, we consider a path over both edges as depicted in Fig.~\ref{fig:Sketch16Cavities}(c). 
Here the signal is split at oscillator 1, travels along the respective edge and interferes at oscillator 16. In a square lattice
both pathways are symmetric and simply correspond to a combination of two one-edge cases discussed in the section above,  i.e., the signal amplitude is
simply doubled at the output oscillator.
Hence, the transmission simply gains a factor of 4, i.e., we have $\mathcal T_{\rm 2 edge}[0] = 4 \mathcal T_{\rm 1 edge}[0]$.

Similarly, the noise added at each path is qualitatively the same. However, instead of seeing a doubling of the added noise, we find half the added noise compared to the one-edge case.
This noise reduction is due to a destructive interference effect, which becomes clear if we consider the noise contribution before each edges first amplification stage.
The latter stages are between oscillator 2 and 3 on the upper edge and at oscillator 5 and 9 on the left edge, thus we need the stationary solutions
\begin{align}\label{Eq.:StatBothEdges}
	 \hat d_2  =&                \mathcal G_2   
	                              \bigg\{  \hat d_{1,\IN}    + i 2 \sqrt{\frac{\Gamma}{\kappa}}  \hat d_{23,\IN}^{\dag}    - i  \sqrt{\frac{1}{2  } }
	                             \left[  \hat d_{67,\IN}   +  \hat d_{610,\IN}  - \hat d_{12,\IN} + \hat d_{15,\IN}  -  \hat d_{26,\IN} +  \hat d_{56,\IN}\right]  \bigg\}, 
  \nonumber \\ 
        \hat d_5=&                   \mathcal G_2 
	 	                     \bigg\{  \hat d_{1,\IN}      + i 2 \sqrt{\frac{\Gamma}{\kappa}}  \hat d_{59,\IN}^{\dag}   - i \sqrt{\frac{1}{2 } }
	                             \left[   \hat d_{67,\IN}   +  \hat d_{610,\IN} + \hat d_{12,\IN} - \hat d_{15,\IN}  +  \hat d_{26,\IN}  - \hat d_{56,\IN}  \right] \bigg\},
\end{align}
at each amplification stage at least $3.5$ quanta is added to the signal (assuming $\widetilde\Gamma \rightarrow \kappa$ and zero temperature), just like in the one-edge case.
However, at oscillator 16 both pathways interfere, this means we have to consider the combination of both contributions as they appear in the stationary solution of oscillator 16,
i.e., $ \hat d_{16} \sim (\hat d_2 + \hat d_5 )/2$. Crucially, the stationary solutions of oscillator 2 and 5 have mainly contributions from the same link-oscillators, cf. Eq.~(\ref{Eq.:StatBothEdges}).
The contributions of the link-oscillators connecting the four oscillators in the left upper corner $(1,2,5 \& 6)$ deconstructly interfere. 
Hence, the noise effectively added after the first two amplification stages can be expressed as
\begin{align}
 \bar n_{25,\rm add}  = \bigg\{ \frac{2 \widetilde\Gamma}{\kappa}  +    1  \bigg\} \left(  \bar n_{\rm link}^T + \frac{1}{2} \right)
                   \rightarrow 3 \left(  \bar n_{\rm link}^T + \frac{1}{2} \right),
\end{align}
this coincides with the  minimum noise value added to a signal leaving the final output port (for large gain). At least 1.5 quanta are added, which is half the single-edge propagation case. This minimum added noise is again independent of the lattice size; it is solely determined by the first amplification stages. 
\subsection*{Various propagation paths}

A crucial aspect of our setup is that we can choose an arbitrary path through the lattice.
We find that, independent of the chosen propagation path and the lattice size, the dynamics at the point of directionality are described by purely real eigenvalues.
Figure \ref{fig:VarPaths} depicts the results for a 64 oscillator lattice and three different path ways.
The case of propagation over all oscillators, i.e., Fig.~\ref{fig:VarPaths}a(i), shows similarities to the cavity chain, 
here the point of directionality coincides with three exceptional points and marks a transition from oscillatory to purely damped dynamics (in the rotated frame).

\section*{Examples for implementations in a superconducting lattice architecture}\label{Sec.:SIimplementation}
%
There are multiple ways to implement the proposed setup as sketched in Fig. 6 in the main text.
In this section we provide further details for these actual implementations. 
The basic building block could for example consists of three cavity modes representing the two node-oscillators and the link-oscillator.
The three modes are coupled via one or more non-linear elements which, under appropriate driving, provides the desired interactions between them.
Crucially, we require a coherent exchange interaction between the node-oscillators, while the coupling to the link-oscillator realizes an indirect exchange interaction. However, instead of using a cavity mode as the link-oscillator one could as well us a highly damped qubit which we discuss in detail in the following section.
\subsection*{A highly damped qubit as an engineered reservoir}
We focus on a single element made out of two node-oscillators with frequencies $\omega_1$ and $\omega_2$, which are described by the operators $\hat d_{1}$ and $\hat d_2$.
We require a coherent hopping interaction of the form
\begin{align}\label{Eq.:HoppHam}
	\hH_{\HOP} =&    M(t) \  \hat d_1^{\dag} \hat d_2 +  h.c.  ,
\end{align} 
where $M(t) = G_{12} \cos(\omega_{P} t + \phi_{P})$ is a time-dependent coupling which is modulated by an external pump. 
Such an interaction can be realized by coupling the two cavities via a superconducting quantum interference
device (SQUID) \cite{peropadre_tunable_2013,yin_catch_2013, pierre_storage_2014}. 
Driving the respective coupling circuit with an external flux at the frequency difference of the cavity modes, i.e., $\omega_{P} = \omega_1 - \omega_2$, induces resonant hopping between them.

The Hamiltonian in Eq.(\ref{Eq.:HoppHam}) realizes reciprocal information transfer between the two cavity modes. 
To break this symmetry we combine this coherent interaction with an engineered dissipative interaction, i.e., we construct an engineered reservoir  
which is connected to the two cavity modes and mediates an effective interaction between them.  
For this we couple both modes to a qubit with transition frequency $\omega_q$.
The qubit itself is an open system, i.e., it is coupled with rate $\gamma$ to an environment which damps its dynamics.
As we will see in what follows, the qubit realizes our engineered reservoir in the high damping case, i.e., when its dynamics is much faster as the one of the cavity modes
and can effectively considered to be a Markovian bath for the cavities.
The combined system is described by the Hamiltonian 
\begin{align} 
 \hH =  \frac{\omega_q}{2} \hat \sigma_z   
      + \sum_{n = 1,2} \left[ \omega_{n} \hat d_{n}^{\dag} \hat d_n + g_n  \left(\hat d_{n} + \hat d_{  n}^{\dag} \right)  \hat \sigma_x  \right] .
\end{align} 
The qubit and the cavity modes interact via a standard Rabi interaction, where $g_n$ corresponds to the coupling between the individual modes and the qubit, 
the latter is described by the Pauli operators $\hat \sigma_{x,y,z}$.   
An external drive at frequency $\Omega$ modulates the transition frequency of the qubit, 
\begin{align}\label{Eq.:QubitDrive}
 \hH_{\DRIVE} =    \Omega \varepsilon \cos( \Omega  t + \phi  ) \hat \sigma_{z}.
\end{align}
a driving scheme which incorporates first order sideband transition physics \cite{porras_shaping_2012,beaudoin_first-order_2012, strand_first-order_2013, navarrete-benlloch_inducing_2014}.
This means that by tuning the drive to a sideband at frequencies $\omega_{q} \pm \omega_{n}$ (blue/red) one can realize first order scattering processes between
the qubit and the cavity modes. 
For example, having a tone on the red sideband results in swapping of excitations between the cavity modes and the qubit, a process which conserves the number of excitations.
On the other side, having a tone on the blue sideband realizes a two-mode squeezing interaction which generates entanglement and amplification.  
A possible realization would be an external flux line in the vicinity of a transmon qubit.

In the next step we introduce an interaction picture with respect to the free Hamiltonian for the cavities and the qubit, as well as to the drive Hamiltonian $\hH_{\DRIVE}$,
described by the unitary transformation  
\begin{align}
 U (t)  =&
            e^{-i    \left[ \omega_1 \hat d_{1}^{\dag} \hat d_1 + \omega_2 \hat d_{2}^{\dag} \hat d_2\right]  t } 
            e^{- \frac{i}{2 }  \omega_q   t}
            e^{  - i \int\limits_0^t dt' \hH_{\DRIVE} (t')}, 
\end{align}
applying this transformation the our system Hamiltonian, i.e.,  $\hH^{\prime} =  U^{\dag} \hH U$, leaves us still with a time-dependent Hamiltonian of the form 
\begin{align}\label{Eq.:HamQubitNonRWA}
 \hH^{\prime} =&         \sum_{n = 1,2}  \hat d_{n}     
              \left[G_{n}^{+}(t) \  \hat \sigma_{-}  
                  + G_{n}^{-}(t) \  \hat \sigma_{+}  \right]  + h.c. , 
 \hspace{0.5cm}
 G_{n}^{\pm}(t) =  \  g_n     \sum_{k  = - \infty}^{+\infty}  \hspace{-0.2cm}
                   J_{k } (2\varepsilon  )     e^{- i  \left[\omega_n   \pm \omega_q \pm k \Omega \right] t    } \ e^{\mp i   k \phi } ,
\end{align}
with the time modulated coupling coefficients $ G_{n}^{\pm}(t) $. So far this Hamiltonian is exact and involves Raman up- and down scattering processes between the cavity resonant frequency and
the sidebands at $\omega_n \pm \omega_q$. However, by choosing the frequency of the external pump one can engineer desired resonant interactions in the coupled system. We choose a special hierarchy of the resonant frequencies involved, setting $\omega_{1} - \omega_q = \omega_{q} - \omega_2 = \Delta$ 
and drive at $\Omega = \Delta$. Then the couplings $G^{-}_n$ become only secular for $k  = \pm 1$, hence, for large enough $\Delta$ we can 
make a rotating wave approximation (RWA) and
 approximate the couplings to $G_{n}^{-}(t) \simeq  \pm g_n J_{1}(2\varepsilon) \ e^{\pm i     \phi_n } \equiv \pm G_n e^{ \pm i     \phi_n }$,
where the $+(-)$ sign refers to $n=1(2)$. Setting $G_1 = G_2 \equiv G$ the effective Hamiltonian becomes
\begin{align}\label{Eq.:HamQubitRWA}
 \hH^{\prime}_{\textup{eff}} =&  \  G  \left(    \hat d_{1} e^{-  i \phi    }  -    \hat d_{2}  e^{- i \phi   } \right) \hat \sigma_{+}  + h.c. . 
\end{align} 
The remaining counter-rotating contributions originating from $G_n^{-}(t)$ are oscillating with $k^{\prime} \Delta$  where $ k^{\prime} \neq 0$
and can be neglected. Additionally, the counter-rotating terms related to the parametric coupling $G_{n}^{+}(t)$ are off resonant for 
\begin{align}
 k_1 \neq& \frac{\omega_1 + \omega_q}{\omega_1 -  \omega_q}, 
 \hspace{0.5cm}
 k_2 \neq  \frac{\omega_2 + \omega_q}{\omega_q -  \omega_2}, 
 \hspace{0.5cm}
 k_{1,2} \in \mathbb{Z},
\end{align}
which can be achieved by appropriate choice of the resonant frequencies. 
 
 \begin{figure*} 
 \centering\includegraphics[width=1.0\textwidth]{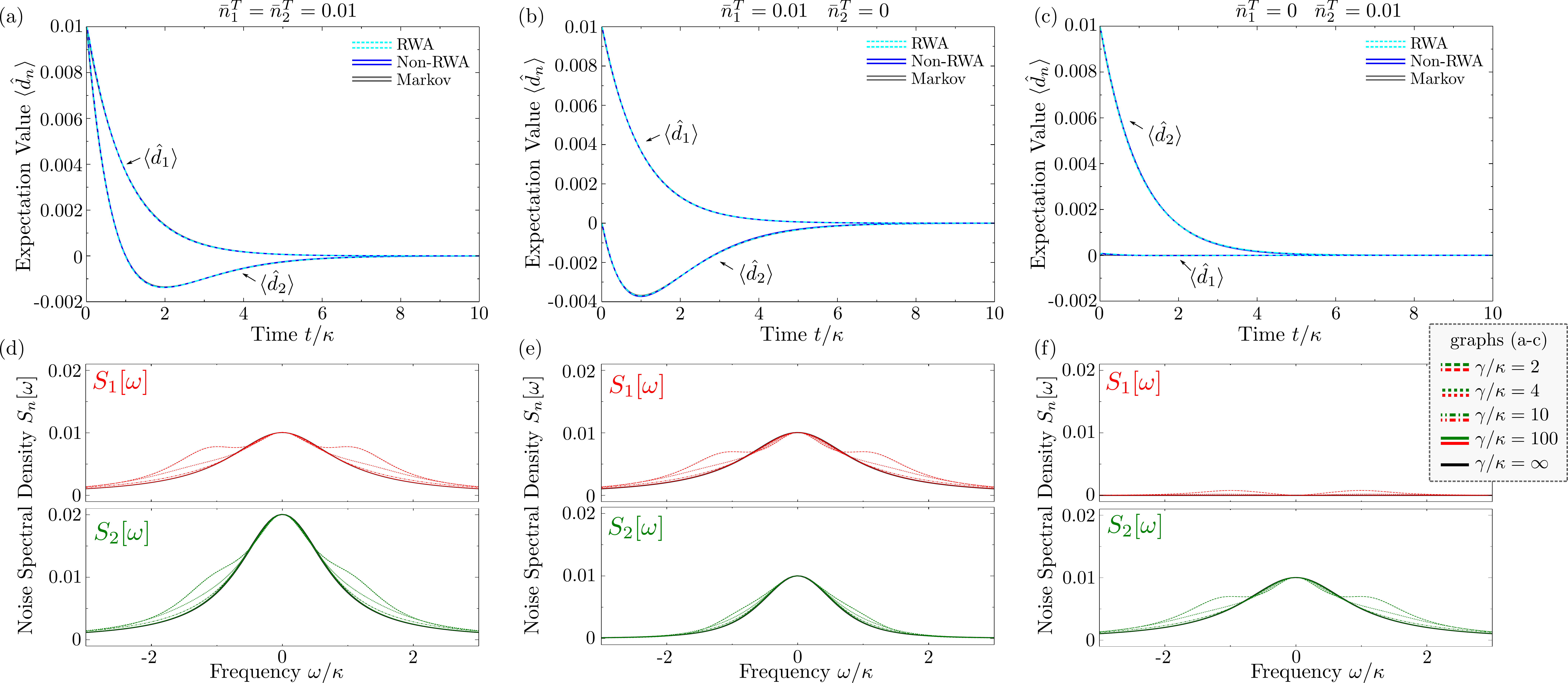} 
	\caption{Numerical simulations of the master equation for the cavities plus qubit system for three different initial conditions.
	        Graph (a-c) depict the dynamics of the cavities expectation value with (Non-RWA) and without (RWA) a rotating wave approximation. 
	        Additionally, the results for the Markovian limit (Markov) are plotted. 
	        Parameters are  $\omega_1/\kappa = 1100$,$\omega_2/\kappa = 100$, $\omega_{q}/\kappa = 600$ and $\gamma/\kappa = 100$. 
	        Graph (d-f) depict the noise spectra for 
	        both cavities for various values of qubit damping rate $\gamma$ (as denoted in the graph).
	        For finite rate $\gamma$ we used a rotating wave approximation,  i.e., we simulated Eq.~(\ref{Eq.:HamQubitRWA}),
	        and compare it to the results for the Markovian master equation (\ref{Eq.:BSmasterequation}).
	        The resulting spectra coincide in the high damping case as expected.}
	\label{fig:DissVSnonRWAvsRWA}
 \end{figure*}

Finally, we include the coherent hopping given in Eq.~(\ref{Eq.:HoppHam}) and
have now our final building block, a two cavity unit where we can tune the interaction direction by adjusting the strength and the phase of
the drive tone on the qubit together with tuning the coherent interaction. An advantage of the chosen frequency hierarchy is, that the pump frequency 
for the coherent interaction is at $\omega_{P} = \omega_{1} -\omega_2 = 2 \Delta $,
which is simply the second harmonic of the drive tone on the qubit. Hence, a single pump source is in principle sufficient.

The qubit is tantamount to a cavity mode as a link-oscillator. To see this, we assume that the qubit is coupled to a zero-temperature bath with decay rate $\gamma$. 
In the case of a highly damped qubit it can be adiabatically eliminated and the system is modeled via the Lindblad master equation
 (setting $\varphi  \equiv \pi - 2 \phi   $)
\begin{align}\label{Eq.:BSmasterequation}
 \frac{d}{dt} \hat \rho =& - i   \left[\hH_{\HOP} , \hat \rho \right] 
			 + \Gamma  \mathcal L [\hat d_1 + e^{i \varphi} \hat d_2 ] \hat{\rho},  
\hspace{0.2cm} 
\Gamma = \frac{4 G^2}{\gamma},
\end{align}
with the superoperator $\mathcal L[\hat o] \hat \rho = \hat o \hat \rho  \hat o^{\dag} - \frac{1}{2} \hat o^{\dag} \hat o \hat \rho - \frac{1}{2} \hat \rho \hat o^{\dag} \hat o$.
This master equation describes the two kinds of interactions between the cavity modes we were aiming for: the coherent hopping $\hH_{\HOP}$ with coupling strength $G_{12}$  
and the dissipative hopping with rate $\Gamma$ assisted by the qubit.

The adiabatic elimination of the qubit is possible in the high damping case; to show that the master equation (\ref{Eq.:BSmasterequation}) describes the system dynamics successfully we 
perform a numerical simulation of the master equation with the full Hamiltonian using Eq.~(\ref{Eq.:HamQubitNonRWA}) and Eq.~(\ref{Eq.:HoppHam}). 
We compare our findings to the RWA solution involving Eq.~(\ref{Eq.:HamQubitRWA}) and the Markovian limit described by Eq.~(\ref{Eq.:BSmasterequation}). 
Figure~\ref{fig:DissVSnonRWAvsRWA}(a-c) depicts the resulting dynamics for the cavities' expectation values $\langle d_{1,2} \rangle =\bar d_{1,2} $ for three different initial conditions
(and assuming always that the directionality conditions are met).
In the Markovian limit their time-dependence is described via (with $\Gamma = \kappa$)
\begin{align}
 \bar d_{1}(t) = \bar d_{1}(0) e^{- \kappa t} ,
 \hspace{0.5cm}
  \bar d_{2}(t) = \left\{ \bar d_{2}(0) - \bar d_{1}(0) \kappa t   \right\} e^{-  \kappa   t} .
\end{align}
The simulated dynamics coincide nicely with these expressions. 
For the case of finite occupation of each cavity at time  $t=0$, i.e.,  $ \bar d_{1}(0) = \bar d_{2}(0)$, 
the expectation values decay fast and $ \bar d_{2}(t)$ becomes negative at $t/\kappa = 1$ as expected.
cf. Fig.~\ref{fig:DissVSnonRWAvsRWA}(a). 
Crucially, for $\bar d_{1}(0) = 0 $ and finite $\bar d_{2}(0) $ the dynamics of cavity 1 is unaffected, 
while for the reversed initial conditions excitations are transfered from cavity 1 to 2,
see Fig.~\ref{fig:DissVSnonRWAvsRWA}(c) and (b) respectively.
 
Figures~\ref{fig:DissVSnonRWAvsRWA}(d-f) depict the resulting noise spectral densities as a function of frequency. 
Here we compare the RWA solution involving Eq.~(\ref{Eq.:HamQubitRWA}) for various values of $\gamma$ and the Markovian limit described by Eq.~(\ref{Eq.:BSmasterequation}).
In the large damping regime the analytical solutions for the noise spectra are ($\bar n^T_{\rm link} = 0$)
\begin{align}
 S_1 [\omega] =&   \frac{1}{\kappa} \frac{ \bar n_{d_1}^{T} }{\left(1  + \frac{\omega^2}{\kappa^2} \right)}           ,
\hspace{0.5cm}	                     
 S_2 [\omega] =    \frac{1}{\kappa}  \left[ \frac{  \bar n_{d_2}^{T}  }{\left( 1  + \frac{\omega^2}{\kappa^2} \right)}    
 	           + \frac{  \bar n_{d_1}^{T} }{\left( 1  + \frac{\omega^2}{\kappa^2}  \right)^2}           \right],
\end{align}
here the cavity-1 spectra has no contribution from cavity 2. These expressions require a large enough damping $\gamma/\kappa$ of the qubit, however, on resonance $(\omega = 0)$  
the system is always nonreciprocal, see Fig.~\ref{fig:DissVSnonRWAvsRWA}(d-f). 
Crucially, the damping $\gamma/\kappa$ determines the frequency range over which one obtains directionality.
%
\subsection*{Implementation with a Josephson Parametric Converter (JPC)}
%
Another possible implementation is based on the Josephson Ring modulator \cite{abdo_nondegenerate_2013_SI,flurin_superconducting_2015},
which is a ring intersected with four Josephson junction realizing three-wave mixing. 
Embedding the latter element into a circuit with three microwave resonators enables quantum limited amplification and conversion of microwave signals;
the complete circuit is called a Josephson Parametric Converter (JPC)\cite{abdo_nondegenerate_2013_SI}. 
Moreover, the JPC can be operated in a nonreciprocal mode, which was recently demonstrated in \cite{sliwa_reconfigurable_2015_SI}. 
In the following we recall the basic ideas of this mode of operation.
The basic system Hamiltonian of the JPC yields
\begin{align}
 \hH = \omega_a \hat a^{\dag} \hat a  + \omega_b \hat b^{\dag} \hat b  + \omega_c \hat c^{\dag} \hat c  
      + g_{3} \left(\hat a + \hat a^{\dag} \right) \left(\hat b + \hat b^{\dag} \right) \left(\hat c + \hat c^{\dag} \right),
\end{align}
where $g_3$ denotes the coupling strength $g_3$ of the three modes with resonant frequencies $\omega_{a,b,c}$. 
We introduce an interaction picture with respect to the free Hamiltonian and obtain 
\begin{align}
 \hH =  g_{3} \left(\hat a e^{-i\omega_{a}t} + \hat a^{\dag} e^{+i\omega_{a}t} \right) 
              \left(\hat b e^{-i\omega_{b}t} + \hat b^{\dag} e^{+i\omega_{b}t}\right) 
              \left(\hat c e^{-i\omega_{c}t} + \hat c^{\dag} e^{+i\omega_{c}t}\right).
\end{align}
Each of the mode is externally driven via an off-resonant pump, the corresponding driving frequencies are $\omega_{P,n}, (n \in a,b,c)$.
In the next step we perform a displacement transformation of the form
\begin{align}
 \hat a = \bar a e^{-i (\omega_{P,a}- \omega_a) t - i \phi_a } + \hat d_{1}  ,
 \hspace{0.5cm}
 \hat b = \bar b e^{-i (\omega_{P,b}-\omega_b) t - i \phi_b } + \hat d_{2}  ,
  \hspace{0.5cm}
 \hat c = \bar c e^{-i (\omega_{P,c}-\omega_c) t - i \phi_c } + \hat d_{3} ,
\end{align}
where we keep the amplitudes  $\bar a, \bar b$ and $\bar c$ real and introduce the pump-phases  $\phi_n$.
The first terms describe the strong field component resulting from the external driving. Note, in this rotated frame we have to subtract the modes resonant frequency from the pump frequency.
We assume strong driving and thus neglect possible fluctuations at the pump frequencies, i.e., we make a stiff pump  approximation.
The second terms describe the field/fluctuations at the  modes' resonant frequencies. 
After some algebra we find for the resonant interactions
\begin{align}
 \hH =& \hspace{0.5cm} g_{3}  \bar c   \bigg[    
                           \hat d_{1}^{\dag} \hat d_{2}^{\dag}  e^{-i( \omega_{P,c} - (\omega_{a}+\omega_b))t - i \phi_c}    
                       +   \hat d_{1}^{\dag} \hat d_{2}   e^{-i(  \omega_{P,c} -(\omega_{a} -\omega_b ))t - i \phi_c} 
                       \bigg] 
                    \nonumber \\ & 
      +  g_{3}  \bar b   \bigg[ 
                           \hat d_{1}^{\dag} \hat d_{3}^{\dag}    e^{-i (\omega_{P,b} -   (\omega_a +\omega_{c}) ) t - i \phi_b }  
                       +   \hat d_{1}^{\dag}  \hat d_{3}    e^{-i (\omega_{P,b} -   (\omega_a - \omega_{c}) ) t - i \phi_b }  \bigg] 
                       \nonumber \\ & 
       +  g_{3}  \bar a   \bigg[                
                             \hat d_{2}^{\dag} \hat d_{3}^{\dag}   e^{-i (\omega_{P,a} - (\omega_b + \omega_{c})) t - i \phi_a } 
                       +     \hat d_{2}^{\dag} \hat d_{3}     e^{-i (\omega_{P,a} - (\omega_b - \omega_{c})) t - i \phi_a }    
                    \bigg]  + h.c. ,
\end{align}
here we made a rotating wave approximation. The choice of the pump frequencies determines the resonant interactions between two modes. 
Pumping at the sum of the frequencies results in non degenerate parametric amplification, i.e., first terms in the upper Hamiltonian.
The second terms describe frequency conversion and are realized if one drives at the frequency difference of two modes. 
The latter would correspond to choosing the pumping scheme
\begin{align}
 \omega_{P,a} =& \; \omega_b - \omega_{c},
 \hspace{0.5cm}
 \omega_{P,b} = \; \omega_a - \omega_{c},
 \hspace{0.5cm}
 \omega_{P,c} =  \; \omega_a - \omega_b,
\end{align}
which results in the effective Hamiltonian
\begin{align}
 \hH_{\textup{int}} =& \hspace{0.5cm}           
            G_{c}  \;   \hat d_{1}^{\dag}  \hat d_{2}   e^{ - i \phi_c}   
       +    G_{b}  \;   \hat d_{1}^{\dag}  \hat d_{3}   e^{ - i \phi_b}   
       +    G_{a}  \;   \hat d_{2}^{\dag}  \hat d_{3}   e^{ - i \phi_a}    + h.c.   ,
       \hspace{0.5cm}
       G_{n} \equiv g_{3}  \bar n 
\end{align}
here we again neglected counter-rotating terms. This is exactly the Hamiltonian for our building block. A basic gauge transformation yields Eq.(1) of the main text. Here one of the modes can be considered as the link-oscillator which realizes the dissipative interaction between the remaining two modes.
We note that one could as well employ a SQUID as the coupling element between the three modes as demonstrated recently \cite{lecocq_nonreciprocal_2016_SI}.
\end{widetext}
%

\end{document}